\documentclass{article}

\usepackage{amssymb,amsfonts,amsmath,stmaryrd}
\usepackage{cite,enumerate,float,indentfirst}
\usepackage{color}
\usepackage{euscript}
\usepackage{tikz}

\DeclareMathOperator\arccosh{arccosh}

\def\be{\begin{eqnarray}}
\def\ee{\end{eqnarray}}
\def\nn{\nonumber}

\def\p{\partial}
\def\tr{{\rm tr}\,}
\def\Tr{{\rm Tr}\,}

\newcommand{\beq}{\begin{equation}}
\newcommand{\eeq}{\end{equation}}
\newcommand{\beqa}{\begin{eqnarray}}
\newcommand{\eeqa}{\end{eqnarray}}

\newcommand{\lam}{\lambda}
\newcommand{\m}{\mu}

\definecolor{red}{rgb}{1,0,0}
\definecolor{orange}{rgb}{1,0.5,0}
\definecolor{violet}{rgb}{0.7,0,1}

\def\cn{{\rm cn}}
\def\sn{{\rm sn}}

\hoffset= - 1.0in         

\begin{document}

\title{\vspace{1.5cm}\bf
On the status of DELL systems
}

\author{
A. Mironov$^{b,c,d,}$\footnote{mironov@lpi.ru,mironov@itep.ru},
A. Morozov$^{a,c,d,}$\footnote{morozov@itep.ru}
}

\date{ }

\maketitle

\vspace{-6.5cm}

\begin{center}
\hfill FIAN/TD-12/23\\
\hfill IITP/TH-14/23\\
\hfill ITEP/TH-20/23\\
\hfill MIPT/TH-15/23
\end{center}

\vspace{4.5cm}

\begin{center}
$^a$ {\small {\it MIPT, Dolgoprudny, 141701, Russia}}\\
$^b$ {\small {\it Lebedev Physics Institute, Moscow 119991, Russia}}\\
$^c$ {\small {\it NRC ``Kurchatov Institute", 123182, Moscow, Russia}}\\
$^d$ {\small {\it Institute for Information Transmission Problems, Moscow 127994, Russia}}
\end{center}

\vspace{.1cm}

\begin{abstract}
A detailed review of the $p,q$-duality for Calogero system and its generalizations is given.
For the first time, we present some of elliptic-trigonometric Hamiltonians dual to the elliptic Ruijsenaars Hamiltonians
(i.e. trigonometric-elliptic ones),
and explain their relations to the bi-elliptic Koroteev-Shakirov (KS) model.
The most interesting self-dual double-elliptic (DELL) system remains a mystery,
but we provide a clearer formulation of the problem and describe the steps that are still to be done.
\end{abstract}

\bigskip

\section{Introduction}

The problem of double-elliptic (DELL) generalization of Calogero-Moser-Sutherland-Ruijsenaars-Schneider
integrable systems \cite{CSM,OPc,OPq,RS,Ruiell,Kri,Cal} remains unresolved.
It was explicitly formulated in \cite{BMMM,MM} as the problem to find a system of commuting
{\it self-dual} bi-elliptic Hamiltonians.
Bi-elliptic means that they have elliptic dependence on coordinates and momenta \cite{BMMMbe}
characterized by two different complex structures (moduli) $\tau$ and $\tilde \tau$.
Self-dual means that they do not change under the Ruijsenaars duality transformation \cite{Ruid} (see
\cite{Etin} for its quantum version).

Examples of {\it bi-elliptic} Hamiltonians were explicitly known from the very beginning \cite{BH},
and now include the beautiful, though straightforward, Koroteev-Shakirov (KS) system \cite{KS}.
Unfortunately, it is still {\it not} self-dual, and does not solve the DELL problem.
The persisting task is to unify the two notions of (self)-duality and (bi)-ellipticity,
which are somewhat different.
Both are difficult, but understood.
Unification is still escaping.

In this paper, we review a current status of the problem.
We explain once again what self-duality means,
and remind the three top achievements in this direction:
the self-dual rational-rational and trigonometric-trigonometric systems,
and the $N=2$ example of elliptic-elliptic system with a highly nontrivial
dependence on $\tau$ and $\tilde\tau$.
Another top achievement is a partial description \cite{MMZ} of the elliptic${_p}$-trigonometric${_q}$
system dual to the well-known trigonometric${_p}$-elliptic${_q}$ Ruijsenaars-Schneider (RS) one:
the wave functions of this system are symmetric polynomials orthogonal to the elliptic polynomials of the generalized Noumi-Shiraishi (GNS) type \cite{GNS}, these latter being eigenfunctions of the degenerate Koroteev-Shakirov Hamiltonians \cite{MMZ}. However, the Hamiltonians dual to the trigonometric${_p}$-elliptic${_q}$ RS ones are still unknown.
A system of eigenfunctions of the full KS Hamiltonians is also known \cite{MMZ}, and is given by the Nekrasov-Shatashvili limit of a peculiar elliptic lift \cite{AKMM2} of the Shiraishi functions \cite{Shi}, however, their transformation to eigenfunctions of self-dual Hamiltonians is unknown. Neither known are the self-dual Hamiltonians themselves.
This sets the stage for a final attack, which still remains to be performed.

We concentrate on the most immediate way to construct DELL systems, though there were a lot more discussion of various DELL system properties \cite{FR,FGNR,GR,BGOR,AMMZ,ABMMZ,AMM,GKKS}.\footnote{See also a discussion of another avatar of the same duality known under the name spectral duality in \cite{specdu}.}

\bigskip

The present situation with dual systems looks as follows.
\begin{itemize}
\item{} Classical duality for two particles in the center of mass frame: one $p$ and one $q$.
It has a reasonably simple formulation and many example can be easily provided, see sec.\ref{N2c}.
The self-dual  ell$_{p}$-ell$_{q}$ case is also known \cite{BMMM} but looks surprisingly complicated.
Even the elliptic modulus becomes dynamical, i.e. $q$-dependent(!).

\item{} Classical duality for $N$ particles is formulated entirely in terms of Hamiltonians and for $N>2$ is rather involved
even for simple systems: in fact, it requires either using the Lax representation \cite{RS}, or, which is equivalent, the Hamiltonian reduction \cite{OPc,HR}. The self-dual ell$_{p}$-ell$_{q}$ case is not done already for $N=3$ (see also \cite{ABMMZ}).

\item{} Quantum duality involves also wave functions, and is {\it formulated} much simpler, roughly as
$$ {\bf H}\left(\frac{\p}{\p{\bf q}}, \hat{\bf q}\right)\Psi(\hat{\bf q}|{\bf \lambda})
= {\bf \lambda}\Psi(\hat{\bf q}|{\bf \lambda})
\Longleftrightarrow
 {\bf H}^D\left(\frac{\p}{\p{\bf \lambda}}, \hat{\bf \lambda}\right)\Psi(\hat{\bf q}|{\bf \lambda})
= {\bf q}\Psi(\hat{\bf q}|{\bf \lambda})
$$
In other words, the quantum means that, if one solved the trig$_{p}$-ell$_{q}$ RS system with a known set of Hamiltonians ${\bf H}$,
one knows the wave functions of the dual ell$_{p}$-trig$_{q}$ one:
these are the $E$'s in sec.\ref{sE}, and it remains ``just" to find a complementary set of operators ${\bf H}^D$ at the r.h.s.
The self-duality means that now there is a set of functions $\Psi(\hat{\bf q}|{\bf \lambda})$,
which is intact under the duality transformation.

\item{} This trick, however, does not allow one directly to deal with quantum {\it self}-duality.
Still, it can be easier to guess self-dual wave functions than Hamiltonians:
look at the trig$_{p}$-trig$_{q}$ case, sec.{\ref{sMac}}.

\item{} A serious assistance is provided by the conjugation procedure
of the $E_\lambda$-polynomials to the KS eigenfunctions $P_\lambda$ in the ell$_{p}$-trig$_{q}$ case:
the latter are also available in the {\it bi}-elliptic (ell$_{p}$-ell$_{q}$) case \cite{MMZ},
and, if the relation between polynomials $E_\lambda$ and $P_\lambda$ persists, one can guess the  ell$_{p}$-ell$_{q}$ version of $E_\lambda$ and then check its self-duality.

\item{} It, however, does not allow one to solve the problem of constructing the corresponding self-dual ell$_{p}$-ell$_{q}$ Hamiltonians, and even their ell$_{p}$-trig$_{q}$ degeneration despite the KS Hamiltonians are known: the problem here is that the relation between $E$ and $P$ as symmetric polynomials of $e^{q_i}$ is immediately realized at the level of Hamiltonians in terms of power sums $\mathfrak{p}_k=\sum_ie^{kq_i}$, and not that immediately in terms of $e^{q_i}$ themselves, sec.\ref{spk}.

\end{itemize}

The celebrated table of dualities in the Calogero-Ruijsenaars family looks as follows:

\bigskip

\begin{tikzpicture}
\draw (0,0) rectangle (16.8,8);
\draw (0,4) -- (16.8,4);
\draw (0,2) -- (16.8,2);
\draw (0,6) -- (16.8,6);
\draw (4.2,0) -- (4.2,8);
\draw (8.4,0) -- (8.4,8);
\draw (12.6,0) -- (12.6,8);
\draw (0,8) -- (4.2,6);
\draw (3,7.3) node{momentum};
\draw (1.3,6.5) node{coordinate};
\draw (2.1,5) node{rational};
\draw (2.1,3) node{trigonometric};
\draw (2.1,1) node{elliptic};
\draw (6.2,7) node{rational};
\draw (10.5,7) node{trigonometric};
\draw (14.6,7) node{elliptic};
\draw (6.3,5) node{rational Calogero};
\draw (10.5,5) node{rational RS};
\draw (14.6,5.3) node{dual elliptic};
\draw (14.6,4.7) node{Calogero-Moser};
\draw (6.3,3.3) node{trigonometric};
\draw (6.3,2.7) node{Calogero-Sutherland};
\draw (10.5,3) node{trigonometric RS};
\draw (14.6,3) node{dual elliptic RS};
\draw (6.3,1) node{elliptic Calogero-Moser};
\draw (10.5,1) node{elliptic RS};
\draw (14.6,1) node{DELL};
\draw[thick,<->] (7,3) parabola (10,4.7);
\draw[thick,<->] (7,1.5) parabola (14,4.5);
\draw[thick,<->] (11.5,1.3) parabola (14,2.5);
\draw[thick,->] (7.8,5.8) arc (100:-170:.3cm);
\draw[thick,->] (12,2.8) arc (100:-170:.3cm);
\draw[thick,->] (16,1) arc (100:-170:.3cm);
\end{tikzpicture}

\bigskip

Here the systems with three possible dependencies on coordinates and on momenta: rational, trigonometric and elliptic are listed, and the dual systems are marked by arrows. The self-dual systems are located on the diagonal of this table.

\bigskip

The eigenfunctions of Hamiltonians of the systems in the second line of the table are all polynomials: the Jack polynomials, the Macdonald polynomials and the $E_\lambda$ polynomials of sec.\ref{El} (those in the first line are just their degenerate)! However, those in the third line are no longer polynomials, moreover, eigenfunctions of the Hamiltonians of the systems in the first two boxes in this line (the elliptic Calogero-Moser and RS models) are better to obtain as a limit from less degenerate power series proposed by J. Shiraishi \cite{Shi}, which is associated with non-stationary elliptic (Calogero-Moser and RS) systems. On the physical side, these Shiraishi series are associated with (Seiberg-Witten) supersymmetric gauge theories (in $4d$ and $5d$) with adjoint matter and codimension two defects, and the limit to the stationary problem eigenfunctions is just the Nekrasov-Shatashvili limit.

The paper is organized as follows. In section 2, we introduce the notion of duality both in the quantum and classical cases. Then we consider a set of examples of dual systems: in section 3, we discuss the simplest example of two-particle classical systems; then, in sections 4 and 5, we consider the example of two-particle quantum systems, and deal with the $N$-particle quantum duality in section 6. Ambiguities of the definition of the quantum duality problem are discussed in section 7. In section 8, we reformulate Hamiltonians of section 6 in terms of $\mathfrak{p}_k$-variables. Section 9 contains some additional comments.

\paragraph{The notation. Symmetric functions.} We use the notation $S_R$ for the Schur function: it is a symmetric polynomial of variables $x_i$ denoted as $S_R(x_i)$, and it is a graded polynomial of power sums $\mathfrak{p}_k=\sum_ix_i^k$ denoted as $S_R\{\mathfrak{p}_k\}$. Here $R$ denotes the partition (or the Young diagram): $R_1\ge R_2\ge\ldots\ge R_{l_R}$, $|R|:=\sum_i^{l_R}R_i$, where $l_R$ is the number of parts of $R$. We also use the notation $S_{R/Q}$ for the skew Schur functions.

Similarly, we denote through $J_R$ the Jack polynomials, and through $M_R$ the Macdonald polynomials \cite{Mac}.
Note that what is usually denoted as parameters $q$ and $t$ in the texts involving the Macdonald polynomials, we denote (secs.6-7) through $e^{-i\hbar}$ and $e^g$ in order to make our presentation closer to first sections of the paper.

\paragraph{The notation. Elliptic functions.} The definitions and properties of various elliptic functions, Jacobi functions and elliptic integrals used in this paper can be found in \cite{BE}. We use the notation $\theta_1(x;\tau)=\tau_1(x)$ for the standard odd $\theta$-function. We also use the following redefined $\theta$-functions:
\be
\theta(2\pi ix)&=&\theta_1(x,\tau)\nn\\
\theta_\tau(x)&=&\prod_{n=0}\Big(1-e^{2\pi i(n\tau+x)}\Big)\Big(1-e^{2\pi i((n+1)\tau-x)}\Big)\sim e^{\pi ix}\theta_1(x,\tau)
\ee
and we use the notation $\mathfrak{q}$ for the elliptic node: $\mathfrak{q}=e^{2\pi i\tau}$.

The elliptic $\Gamma$-function is defined to be
\be\label{EG}
\Gamma(z;\mathfrak{q},w):={(\mathfrak{q}w/z;\mathfrak{q},w)_\infty\over (z;\mathfrak{q},w)_\infty}=\exp\left[\sum_m{z^m-(\mathfrak{q}w/z)^m\over (1-\mathfrak{q}^m)(1-w^m)m}\right]
\ee
with the Pochhammer symbol
\be\label{Poch}
(z;w_1,\ldots,w_n)_\infty:=\prod_{k_1,\ldots,k_n=0}^\infty (1-zw_i^{k_i})
\ee
The elliptic Pochhammer symbol is defined as
\beq\label{Theta}
\Theta(z;\mathfrak{q},w)_n: ={\Gamma(\mathfrak{q}^nz;\mathfrak{q},w)\over\Gamma(z;\mathfrak{q},w)}
\eeq

\section{Duality map}

\subsection{Duality map in a quantum system}

Consider an integrable system with a set of commuting quantum Hamiltonians $\hat H_i(\hat p,\hat q;g)$ (i.e. with a prescribed ordering of operators) with the property $\hat H_i(\hat p,\hat q;0)=\hat H_i^{(0)}(\hat p)$. Duality is defined to be a map from the set of Hamiltonians $\hat H_i(\hat p,\hat q|g)$ to a set of dual Hamiltonians $\hat H^D_i(\hat p,\hat q|g)$ given as follows.

Consider the set of eigenvalue equations
\be\label{qduality}
\hat H_i(-i\hbar\p_{q_i},q_i|g)\Psi(\lambda_i,q_i)&=&H_i^{(0)}(\lambda_i)\Psi(\lambda_i,q_i)
\ee
where $H_i^{(0)}(\lambda_i)$ are now functions of a variable $\lambda$ (symbols of operators $\hat H_i(\hat p_i,\hat q_i|0)$).

Then, a set of dual Hamiltonians is defined to be
\be\label{qdual}
\hat H_i^D(-i\hbar\p_{\lambda_i},\lambda_i|g)\Psi(\lambda_i,q_i)&=&H_i^{D,(0)}(q_i)\Psi(\lambda_i,q_i)
\ee
where $H_i^{D,(0)}(q_i)$ are functions that can be chosen arbitrarily, since any functions of dual Hamiltonians are still Hamiltonians. However, they have to be chosen in such a way that $\hat H_i^D$ are well-defined (non-singular) operators.

There is freedom in this definition, which will be more clear in the quasiclassical limit.

\subsection{Quasiclassical limit}

As a first step, we consider the quasiclassical limit of this system. In order to save notations, we first deal with a system with one degree of freedom and one non-trivial Hamiltonian. The wave function for the problem (\ref{qduality}) in the quasiclassical limit is
\be\label{wf1}
\Psi_Q(q)=\exp\left({i\over\hbar}\int^q d\tilde qp(\tilde q,H^{(0)}(Q))\right)
\ee
where $p(q,H^{(0)}(Q))$ is a solution to the equation
\be\label{cd1}
H(p,q|g)&=&H^{(0)}(Q)
\ee
Indeed, in the leading order in $\hbar$, one obtains
\be
H(-i\hbar\p_q,q)\Psi_Q(q)=H(p(q,H^{(0)}(Q)),q)\Psi_Q(q)=H^{(0)}(Q)\Psi_Q(q)
\ee
which is the eigenvalue problem for the Hamiltonian $H(-i\hbar\p_q,q)$ with the eigenvalue (energy) $H^{(0)}(Q)$.
We denoted here the dual (spectral) variable via $Q$ instead of $\lambda$ for a reason that will be clear in a moment.

Similarly, the dual problem (\ref{qdual}) is described by the wave function
\be\label{wf2}
\Psi_Q(q)=\exp\left({i\over\hbar}\int^Q d\tilde QP(\tilde Q,H^{D,(0)}(q))\right)
\ee
where $P(Q,H^{D,(0)}(q))$ is the solution to equation
\be\label{cd2}
H^D(P,Q|g)&=&H^{D,(0)}(q)
\ee
However, (\ref{wf1}) and (\ref{wf2}) describe the same wave function. Hence, there is an equality
\be\label{S}
\int^q d\tilde qp(\tilde q,H^{(0)}(Q))=\int^Q d\tilde QP(\tilde Q,H^{D,(0)}(q))
\ee
or, differentiating it in $q$ and $Q$,
\be\label{cd3}
{\p p(q,H^{(0)}(Q))\over\p Q}={\p P(Q,H^{D,(0)}(q))\over\p q}
\ee
This means that we make an anti-canonical transformation (i.e. $dp\wedge dq=-dP\wedge dQ$) from variables $(p,q)$ to variables $(P,Q)$, with a generating function $S(q,Q)$ of this anti-canonical transformation
\be
S(q,Q):=\int^q d\tilde qp(\tilde q,H^{(0)}(Q))=\int^Q d\tilde QP(\tilde Q,H^{D,(0)}(q))
\ee
such that
\be
{\p S(q,Q)\over\p Q}=P,\ \ \ \ \ {\p S(q,Q)\over\p q}=p
\ee

\subsection{Classical duality map}

Having the quasiclassical limit of the duality map, we can formulate the duality map for the classical system without any references to the quantum one, just basing on Eqs.(\ref{cd1}), (\ref{cd2}) and (\ref{cd3}).
That is, consider an integrable system with a set of Hamiltonians $H_i(p,q;g)$ with the property $H_i(p,q;0)=H_i^{(0)}(p)$. Duality is defined to be a map from this set of Hamiltonians $H_i(p,q|g)$ to a set of dual Hamiltonians $H^D_i(p,q|g)$ given by the following anti-canonical change of variables
\be\label{duality}
\left\{
\begin{array}{rcl}
H_i(p_i,q_i|g)&=&H_i^{(0)}(Q_i)\cr
&&\cr
\sum_idp_i\wedge dq_i&=&-\sum_idP_i\wedge dQ_i
\end{array}
\right.
\ee
There is a freedom in this definition, e.g., one can add to $P_i$ arbitrary functions of $Q_i$.

Then, the set of dual Hamiltonians is defined to be
\be\label{dualH}
H_i^D(P_i,Q_i|g)&=&H_i^{D,(0)}(q_i)
\ee
It is not much of importance how to choose $H_i^{D,(0)}$ since a function of Hamiltonians is still a Hamiltonian, however, one has to choose it in such a way that $H_i^D$ is a well-defined (single-valued) function on the symplectic manifold.

\section{Classical duality at $N=2$\label{N2c}}

Now we start considering examples of dual systems, and begin with a set of simple two-particle examples with interaction depending only on the differences of coordinates and in the center-of-mass frame, i.e. we choose the total momentum to be zero, $p_1+p_2=0$, and denote $p_1=p=-p_2$, $q=q_1-q_2$. This system has one non-trivial degree of freedom, one can look at it as a one-particle system in an external field equally well. Most of these examples have been considered earlier \cite{BMMM,MMpq,Zak}. Note that, in this case, the dual system is algorithmically constructed: one derives from the conditions
\be\label{N21}
H(p,q)=H^{(0)}(Q)\nn\\
H^D(P,Q)=H^{D,(0)}(q)\nn\\
dp\wedge dq=-dP\wedge dQ
\ee
the equation
\be\label{N2main}
{\p H^{(0)}(Q)\over \p Q}\cdot {\p H^D(P,Q;g)\over \p P}={\p H^{D,(0)}(q)\over \p q}\cdot {\p H(p,q;g)\over \p p}
\ee
which, along with the first two lines of (\ref{N21}), gives rise to the dual Hamiltonian $H^D(P,Q;g)$.

\subsection{A set of simple examples}

\paragraph{Harmonic oscillator.} The Hamiltonian is
\be\label{ho}
H(p,q|g) = \frac{1}{2}(p^2 + g^2q^2)
\ee
Choosing $H^{D,(0)}=q$, one obtains
\be\label{hod}
H^D(P,Q|g) =
\frac{Q}{g}\sin \frac{g P}{Q}
\ee

\paragraph{Non-periodic Toda chain (Liouville equation).} The Hamiltonian is
\be\label{HL}
H(p,q)={p^2\over 2}+ge^q
\ee
and we choose $H^{D,(0)}=e^q$. Then,
\be
H^D(P,Q)={Q^2\over 2g\cosh^2 P/2}
\ee

\paragraph{Linear potential.} In the case of Hamiltonian
\be\label{HLin}
H(p,q)={p^2\over 2}+g|q|
\ee
we choose $H^{D,(0)}=q$. Then,
\be
H^D(P,Q)=\xi\left({g|\xi|\over 4Q^2}-1\right),\ \ \ \ \ \ \xi:={2Q^2\over g}-P
\ee

\paragraph{Coulomb potential.} In this case, the Hamiltonian is
\be\label{HC}
H(p,q)={p^2\over 2}-{g\over q}
\ee
and we choose $H^{D,(0)}=q$. Then, the dual Hamiltonian is determined as a solution $H^D(P,Q)=x$ to the equation
\be
Qx+\xi(x)+{g\over 2Q}=\exp\left\{2g^{-1}\Big(\xi(x)-Q^2P\Big)\right\}
\ee
where
\be
\xi(x):=\sqrt{x(Q^2x+g)}
\ee

\subsection{Trigonometric RS model}

In all examples of the previous subsection, one can see that the dual Hamiltonian looks completely different from the original one, and the systems are far from self-dual ones. Now we consider an example of the self-dual system: the trigonometric RS model. Note that it is not that unique self-dual system, there are many more \cite{MMpq}. However, systems of the Calogero-Ruijsenaars family are of special interest, since their duality is immediately extended to $N$-particle case, and it has interpretations in various terms \cite{Ruid,FR,FGNR,MM,GM,AMM}.

Thus, we consider the trigonometric RS system and two-particle case, $N=2$ in the centre-of-mass frame, and the dynamics is given by the Hamiltonian
\be\label{tRS}
H_{tR}(p,q;g)={\sinh(q+g)\over 2\sinh(q)}e^p+{\sinh(q-g)\over 2\sinh(q)}e^{-p}
\ee
Then, one can choose $H^{D,(0)}=q$ and obtain for the dual Hamiltonian
\be\label{tRSm}
H^D(P,Q|g)=\arccosh\left({\sqrt{\sinh(Q+g)\sinh(Q-g)}\over\sinh Q}\cosh P\right)
\ee
This immediately implies that a clever choice would be $H^{D,(0)}=\cosh q$ so that
\be\label{tRS2}
H^D(P,Q|g)={\sqrt{\sinh(Q+g)\sinh(Q-g)}\over\sinh Q}\cosh P
\ee
Now, let us note that one is free to shift the momentum $p$ with an arbitrary function of the coordinate $q$. In particular, the shift $P\to P+{1\over 2}\log {\sinh(Q+g)\over\sinh(Q-g)}$ leads to the Hamiltonian (\ref{tRS})
\be
H^D(P,Q;g)={\sinh(Q+g)\over 2\sinh(Q)}e^P+{\sinh(Q-g)\over 2\sinh(Q)}e^{-P}
\ee
i.e. the system is self-dual.

In fact, other dualities in this Calogero-Ruijsenaars family can be described by degeneration of the trigonometric RS system. That is, degenerating the trigonometric dependence of the RS Hamiltonian (\ref{tRS2}) on both coordinate and momentum to the rational one, gives rise to the self-dual rational Calogero model with the Hamiltonian
\be\label{rC}
H_{rC}(p,q;g)={p^2\over 2}-{g^2\over 2q^2}
\ee
At the same time, degenerating the trigonometric dependence on the momentum only gives rise to trigonometric Calogero-Sutherland (CS) Hamiltonian
\be\label{tC}
H_{tC}(p,q;g)={p^2\over 2}-{g^2\over 2\sinh^2 q}
\ee
which is dual to the rational RS Hamiltonian
\be\label{rR}
H_{rR}(p,q;g)=\sqrt{1-{g^2\over q^2}}\cosh p
\ee
describing degenerating the trigonometric dependence on the coordinate only.

\subsection{Periodic Toda chain}

Another example of the same Calogero-Ruijsenaars family is the periodic Toda chain, which can be obtained from the elliptic Calogero model in the Inozemtsev limit \cite{Ino}.

In this case, the Hamiltonian is
\be\label{HT}
H(p,q)={p^2\over 2}+{g^2\over 2}\cosh^2q
\ee
and we choose $H^{D,(0)}=\cosh q$. Then,
\be
H^D(P,Q)=\sn \Big(iP\Big|gQ^{-1}\Big)
\ee
Now one can use the sum formula
\be
\sn (u|k)={\pi\over 2kK}\sum_{n\in\mathbb{Z}+1/2}{\sin n\pi u/K\over\sin n\pi\tau}
\ee
in order to obtain
\be\label{dTC}
H^D(P,Q)={\pi Q\over 2gK}\sum_{n\in\mathbb{Z}+1/2}{\sinh {n\pi P\over K}\over\sinh{n\pi K'\over K}}
\ee
where $K$ is the complete elliptic integrals of the first kind, $K=F\Big(1\Big|{g\over Q}\Big)$, and $K'=F\Big(1\Big|\sqrt{1-{g^2\over Q^2}}\Big)$.

\subsection{Dual elliptic RS model}

Already the periodic Toda chain reveals some ellipticity: the dual Hamiltonian is an elliptic function with the dynamical elliptic modulus depending on the coordinate. Now we come to the system elliptic from the very beginning: we consider the elliptic RS system and again the two-particle $N=2$ case. In this case, the Hamiltonian is
\be\label{eRS}
H(p,q;g)={\theta_1(q+g)\over\theta_1(q)}e^p+{\theta_1(q-g)\over\theta_1(q)}e^{-p}
\ee
and one obtains for the dual Hamiltonian:
\be
\int^{H^D} dH^{D,(0)}(q)\sinh Q
\left( {\p H^{D,(0)}(q)\over \p q}\sqrt{\cosh^2Q-{\theta_1(q+g)\theta_1(q-g)\over\theta_1^2(q)}}\right)^{-1}=P+f(Q)
\ee
where $f$ is an arbitrary function. Choosing, at the moment, $H^{D,(0)}(q)=q$, one gets
\be
\int^{H^D} {\sinh Q dq\over \sqrt{\cosh^2Q-\theta_1^2(g)(\wp(g)-\wp(q))}}=P+f(Q)
\ee
Making a substitution $\wp(q)=x$, one obtains for this integral
\be
\int^{\wp(H^D)} {dx\over 2\sqrt{(\alpha+\beta^2x)(x-e_1)(x-e_2)(x-e_3)}}=P+f(Q)
\ee
where $\alpha=1+(1-\theta_1^2(g)\wp(g))\sinh^{-2}Q$, $\beta=\theta_1 (g)\sinh^{-1}Q$, and $e_i$ are the roots of the equation for the Weierstrasse function $\wp(z)'^2=4(\wp(z)-e_1)(\wp(z)-e_2)(\wp(z)-e_3)$.

This integral can be evaluated at various choices of contours, which corresponds choosing different anti-canonical transformations. W choose
\be\label{Ic}
P+f(Q)=-{1\over\sqrt{(\alpha+\beta^2 e_2)(e_1-e_3)}}F\left(i\sqrt{(\alpha+\beta^2\wp(H^D))(e_1-e_3)\over
(e_3-\wp(H^D))(\alpha+\beta^2 e_1)},\sqrt{{e_2-e_3\over e_1-e_3}\cdot{\alpha+\beta^2e_1\over\alpha+\beta^2e_2}}\right)
\ee
where $F(x,k)$ is the elliptic integral of the first kind.

Now one can use the definition of the elliptic Jacobi functions: if
\be
F(w,k)=u
\ee
then the elliptic Jacobi sine function is defined to be
\be
w=\sn (u|k)
\ee
Hence,
\be\label{1}
i\sqrt{(\alpha+\beta^2\wp(H^D))(e_1-e_3)\over
(e_3-\wp(H^D))(\alpha+\beta^2 e_1)}=\sn \left(-\sqrt{(\alpha+\beta^2 e_2)(e_1-e_3)}P\left|\sqrt{{e_2-e_3\over e_1-e_3}\cdot{\alpha+\beta^2e_1\over\alpha+\beta^2e_2}}\right)\right.
\ee
and we keep in mind that the momentum $P$ can be shifted by an arbitrary function of $Q$.

Now let us use that $\sn^2(u|k)+\cn^2(u|k)=1$ and that
\be
\cn(u|k)=\sqrt{\wp(z)-e_1\over\wp(z)-e_3}
\ee
where $u=\sqrt{e_1-e_3}z$. Then, from (\ref{1}), one obtains
\be
\cn (\sqrt{e_1-e_3}H^D|k)=\sqrt{\alpha+\beta^2e_1\over\alpha+\beta^2e_3}\cn\left(\sqrt{(\alpha+\beta^2 e_2)(e_1-e_3)}P\left|\sqrt{{e_2-e_3\over e_1-e_3}\cdot{\alpha+\beta^2e_1\over\alpha+\beta^2e_2}}\right)\right.
\ee
From this expression, this is clear that the most clever choice of $H^{D,(0)}(q)$ is given not by $H^{D,(0)}(q)=q$, but by $H^{D,(0)}(q)=\cn (\sqrt{e_1-e_3}q|k)$, the final result being
\be
H^D(P,Q|k) =\sqrt{\alpha+\beta^2e_1\over\alpha+\beta^2e_3}\cn\left(\sqrt{(\alpha+\beta^2 e_2)(e_1-e_3)}P\left|k\sqrt{{\alpha+\beta^2e_1\over\alpha+\beta^2e_2}}\right)\right.
\ee
where we used that the elliptic modulus
\be
k^2={e_2-e_3\over e_1-e_3}
\ee
In the degenerate limit $e_1=-2/3$, $e_2=e_3=1/3$, one obtains
\be
H^D(P,Q)=\sqrt{1-{\sinh^2g\over\sinh^2Q}}\cosh P
\ee
which returns us back to (\ref{tRS2}).

With another choice of the integration contour in (\ref{Ic}), one can get the answer \cite{BMMM}\footnote{In \cite{BMMM}, the starting Hamiltonian was
\be\label{edRS1}
H(p,q)=\sqrt{1-\frac{\sn^2g}{\sn^2q}}\cdot \cosh p\nn
\ee
which is obtained from (\ref{eRS}) by the shift $p\to p-{1\over 2}\log {\theta_1(q+g)\over\theta_1(q-g)}$,
\be
H(p,q,|g)=\sqrt{\theta_1(q+g)\theta_1(q-g)\over\theta_1^2(q)}\cosh{p}\nn
\ee
generalizing formula (\ref{tRSm}), and by further rescaling the coordinate $q\to(e_1-e_3)^{-1/2}q$ with a similar rescaling of $g$.
}
\be\label{deRS2}
H^D(P,Q|k) = \alpha_Q\cdot
\cn\left(\sqrt{(k'^2+k^2\alpha^2_Q)(e_1-e_3)}P\ \left|\
\frac{k\alpha_Q}{\sqrt{k'^2+k^2\alpha^2_Q}}  \right.\right)
\ee
where
\be
\alpha^2_Q = 1-\frac{\sinh g^2}{\sinh^2Q}={\sqrt{\sinh(Q+g)\sinh(Q-g)}\over\sinh Q}
\ee
Here again one can shift $p$ by an arbitrary function of $q$. In the limit of $k\to 0$, this Hamiltonian gives rise to (\ref{tRS}).

Let us rewrite this result in the series form:
\be\label{deRS}
H(P,Q)=\alpha_Q\cdot {\pi\over k_QK_Q}\sum_{n\in \mathbb{Z}}{1\over \mathfrak{q}_Q^{n-1/2}+\mathfrak{q}_Q^{-n+1/2}}\cos {\pi u(n-1/2)\over K_Q}
\ee
where we denoted through $\mathfrak{q}$ the elliptic node, and
\be
u=\sqrt{e_1-e_3}\sqrt{k'^2+k^2\alpha_Q^2}P,\ \ \ \ \ k_Q={k\alpha\over\sqrt{k'^2+k^2\alpha_Q^2}}
\ee
The question is {\bf what is a clever choice of an arbitrary shift of momentum that would allow one to provide a proper quantum Hamiltonian.} For instance, one can easily get
\be
H(P,Q)=\alpha_Q\cdot {\pi\over k_QK_Q}\sum_{n\in \mathbb{Z}}\left(
{\exp\left({i\pi u\over K_Q}(n-1/2)\right)\over 1+\mathfrak{q}_Q^{1-2n}}+
{\exp\left(-{i\pi u\over K_Q}(n-1/2)\right)\over 1+\mathfrak{q}_Q^{2n-1}}\right)
\ee
In the limit of $k\to 0$, it leads to (\ref{tRS2}) instead of (\ref{tRS}).

\subsection{Double-elliptic system}

At last, we construct the double-elliptic Hamiltonian, which is self-dual. It is achieved by elliptizing the $q$-dependence of the Hamiltonian of the previous section. As we saw, there are a few Hamiltonians depending on the choice of the integration contour. We choose the Hamiltonian (\ref{deRS2}), and its double-elliptic version is given by just the same Hamiltonian \cite{BMMM}, but with the substitution
\be
\alpha_Q\longrightarrow\alpha_Q^{de}=\sqrt{1-\frac{\sn^2g}{\sn^2q}}
\ee
as in the first formula in footnote 2.

\section{Quantum duality at $N=2$: a set of simple examples\label{qdN2}}

Now we also demonstrate several examples of the quantum duality in $N=2$ case.

\paragraph{Harmonic oscillator.}
The wave functions of the quantum harmonic oscillator with Hamiltonian (\ref{ho}) are the Hermite polynomials, which can be written in the form \cite{LL}
\be\label{howf}
\psi_Q(q) = \xi_n(q) :={(-1)^n\over\sqrt{2^nn!}}e^{z^2/2}{d\over dz}e^{-z^2},\ \ \ \ \ \ z:=\sqrt{\hbar g}\ q
\ee
where the Hamiltonian eigenvalues are\footnote{In the quasiclassical approximation, the r.h.s. of this equation is just $Q^2/2$, how it should be.}
\be\label{ev}
\Big(n+{1\over 2}\Big)\hbar g = \frac{Q^2+\hbar g}{2}
\ee
Since the oscillator Hamiltonian has a discrete spectrum, the dual Hamiltonian should provide
a finite-difference equation, i.e. depend exponentially on $P$. Indeed, the wave functions (\ref{howf}) satisfy the
finite-difference equation
\be
\sqrt{2n+2}\ \xi_{n+1}+\sqrt{2n}\ \xi_{n-1}=2z\xi_n
\ee
Note that the normalization of the wave functions in the duality problem has to be properly chosen. Let us multiply it by $i^n$: $\xi_n\to i^n\xi_n$. Then, using (\ref{ev}), one get the equation
\be
\sqrt{Q^2+2\hbar g}\ \xi_{n+1}-Q\ \xi_{n-1}=2zi\xi_n
\ee
Since from (\ref{ev}) it is clear that the exponential $e^{2g\frac{\p}{\p Q^2}}$ just makes a shift $n\to n+1$ in $\xi_n(q)$, the dual Hamiltonian in this case is given by the operator
\be
\hat{H}^D =
\frac{e^{2\hbar g\frac{\p}{\p Q^2}}Q - Qe^{-2\hbar g\frac{\p}{\p Q^2}}}{2i\hbar g}
\ee
In the quasiclassical approximation, this operator reduces to formula (\ref{hod}).

 \paragraph{Self-dual rational Calogero model.} For the sake of brevity, from now on, we put $\hbar=1$.
An example of self-dual Hamiltonian, which is obtained from the self-dual trigonometric RS system in the rational limit, is the rational Calogero Hamiltonian (\ref{rC}).
The quantum Sch\"odinger equation is
\be\label{rrC}
\left[-{\p^2\over\p q^2}-{g^2\over 2q^2}\right]\psi_Q(q)=Q^2\psi_Q(q)
\ee
Solution to this equation is
\be
\psi_Q(q)=\sqrt{q}J_\nu(Q q),\ \ \ \ \ \ \nu=\sqrt{1/4-g^2/2}
\ee
where $J$ is either of the Bessel functions of the first kind. Now we multiply the wave function with $\sqrt{Q}$:
$\psi_Q(q)\to\sqrt{Q}\psi_Q(q)$. It evidently satisfies the same equation (\ref{rrC}) w.r.t. the variable $Q$:
\be
\left[-{\p^2\over\p Q^2}-{g^2\over 2Q^2}\right]\psi_Q(q)=q^2\psi_Q(q)
\ee
which means self-duality.

In this case, we have the scaling symmetry, no discrete spectrum, and a peculiar realization
of the self-duality: the wave function depends just on the product $Q q$.

\paragraph{Trigonometric CS vs rational RS models.}
There is a degeneration of the trigonometric RS system to the rational dependence on momentum: it is the trigonometric CS model with the Hamiltonian (\ref{tC}).
This model is not self-dual: the dual Hamiltonian is the degeneration of the trigonometric RS Hamiltonian to the rational dependence on coordinate, it is called the rational RS model. In order to construct the dual Hamiltonian, note that the equation for the wave function for the CS Hamiltonian (\ref{tC}),
\be\label{qCS}
\left[-{\p^2\over\p q^2}-{g^2\over \sinh^2q}\right]\psi_Q(q)=Q^2\psi_Q(q)
\ee
has polynomial solutions
\be\label{Jack}
\psi_Q(q)=\sinh^\beta q\cdot J_{[k]}^{(\beta)}(e^q,e^{-q})=\sinh^\beta q\cdot \sum_{l=0}^k e^{(k-2l)q}\prod_{i=0}^{l-1}{(k-i)(\beta+i)\over (i+1)(k-i-1+\beta)},\ \ \ \ \
g^2=-\beta(\beta-1)
\ee
Here $J_{[k]}^{(\beta)}(x_1,x_2)$ is the Jack symmetric polynomial of two variables $x_1$ and $x_2$, and the eigenvalue is
\be
Q^2=-(\beta+k)^2
\ee
The Jack polynomial $J_{[k]}^{(\beta)}(e^q,e^{-q})$ satisfies the three-term recurrent relation in the index $k$:
\be
J_{[k+1]}^{(\beta)}(e^q,e^{-q})+{k(2\beta+k-1)\over(\beta+k)(\beta+k-1)} J_{[k-1]}^{(\beta)}(e^q,e^{-q})=(e^q+e^{-q})J_{[k]}^{(\beta)}(e^q,e^{-q})
\ee
This equation can be realized as the eigenvalue problem with the Hermitian Hamiltonian
\be
\hat H^D=\exp\left(i{\p\over\p Q}\right)+{Q-i\beta\over Q}\exp\left(-i{\p\over\p Q}\right){Q+i\beta\over Q}
\ee
and
\be
\hat H^D\psi_Q(q)=(e^q+e^{-q})\psi_Q(q)
\ee
In the quasiclassical limit, $\hat H^D$ becomes the rational RS Hamiltonian $H_{rR}(P,Q;g)$ (\ref{rR}) with $g^2=-\beta^2$ upon the shift $P\to P-{1\over 2}\log {Q^2-g^2\over Q^2}$. It is also transformed by a unitary transformation $U$ to
\be
\hat H^d\stackrel{U}{\longrightarrow}{Q+i\beta\over Q}\exp\left(i{\p\over\p Q}\right)+{Q-i\beta\over Q}\exp\left(-i{\p\over\p Q}\right)
\ee
At the classical level, this corresponds to another choice of the variable $P$ (the freedom in the definition of the duality map).

\section{First elliptic example: quantum periodic Toda chain}

Now we are ready to consider a truly non-trivial example of quantum duality of the periodic Toda chain, the first example of elliptic dual Hamiltonian.

Following our general scheme, consider the solution to the Schr\"odinger equations with the Hamiltonian (\ref{HT}). Put $g=1$. Then, a solution to the eigenvalue equation
\be
\left[-{\p^2\over\p q^2}+\cosh^2q\right]\psi_Q(q)=Q^2\cdot\psi_Q(q)
\ee
is \cite{Gutz,KhL}
\be
\psi_Q(q)=\sum_k{\phi_ke^{-2(\beta+k)q}\over \Gamma(1+\beta+k+i\sqrt{Q^2-2})
\Gamma(1+\beta+k-i\sqrt{Q^2-2})}
\ee
where $\beta$ is a function of $Q$, and $\phi_k$ satisfies the recurrent relation
\be
\phi_k=\phi_{k-1}+{\phi_{k+1}\over \Big(Q^2-2+(\beta+k+1)^2\Big)\Big(Q^2-2+(\beta+k)^2\Big)}
\ee
its solution being given by the determinant
\be
\phi_k=\det_{\infty> i,j\ge 1}\left(\delta_{i,j}+{1\over Q^2-2+(\beta+k+i)^2}\delta_{i,j-1}+
{1\over Q^2-2+(\beta+k+i)^2}\delta_{i,j+1}\right)
\ee
and the second independent solution corresponds to $\beta\to -\beta$. In order to relate $\beta$ and $Q$, one has to solve the equation
\be
\det_{\infty> i,j\ge -\infty}\left(\delta_{i,j}+{1\over Q^2-2+(\beta+k+i)^2}\delta_{i,j-1}+
{1\over Q^2-2+(\beta+k+i)^2}\delta_{i,j+1}\right)=0
\ee
The problem is that now the system has no equidistant spectrum, which means there are no finite-term recurrent relations for the wave function: it has infinitely many terms. This does not come as a surprise since the dual Hamiltonian is not trigonometric, but elliptic: (\ref{dTC}) is an infinite series.

\section{Quantum duality at arbitrary $N$}

$N$-particle duality at $N>2$ is more difficult to deal with.
Here we start with an explicitly known example of the trigonometric RS model, consider it in detail,
and then discuss a less understood system dual to the elliptic RS one. In the latter system, we can construct
a set of the dual wave functions (using the KS Hamiltonians at the intermediate stage), but the dual Hamiltonians are still lacking.
Throughout this section, we consider the quantum duality only, the classical counterpart being immediate.
For the sake of convenience, we restore the constant $\hbar$ in this section.

\subsection{Trigonometric RS system\label{sMac}}

Quantum Hamiltonians of the $N$-body trigonometric RS system can be realized as the Macdonald difference operators \cite[p.315, Eq.(3.4)]{Mac}
\begin{equation}
  \label{eq:26}
   \hat  H_k=e^{{(N-k)k\over 2}g}\sum_{
      \begin{smallmatrix}
        I \subset
        \{1,\ldots,N \}\\
        |I|=k
      \end{smallmatrix}
}\left[ \prod_{{m \in \{1,\ldots,N\} \backslash I}\atop{l \in I}}  \frac{\sinh {q_l-q_m+g\over 2}}{\sinh{q_l-q_m\over 2}} \right] \prod_{j \in I} e^{-i\hbar \p_{q_j}}
\end{equation}
We consider another natural choice of the quantum Hamiltonians of the $N$-body trigonometric RS system at the very end of this subsection.

The eigenvalue problem
\be\label{evpM}
\hat  H_k\Psi_{Q}(q_l)=e_k \Big(e^{Q_j}\Big)\Psi_{Q}(q_l)
\ee
where $e_k$ are the elementary symmetric polynomials,
is solved by the Macdonald polynomials,
\be\label{efM}
\Psi_{Q}(q_l)=M_\lambda\Big(e^{q_l};e^{-i\hbar},e^g\Big)
\ee
Here use the standard definition of the Macdonald polynomial as a symmetric polynomial of variables $x_i$ depending on two parameters ${\bf q}$ and ${\bf t}$, $M_\lambda(x_i;{\bf q},{\bf t})$ following the book \cite{Mac}, where ${\bf q}$ should not be associated with a coordinate, it is just a parameter.

In (\ref{efM}), $\lambda$ is the partition with parts $\{\lambda_l\}$ that parameterizes the wave function and the eigenvalue so that
\be\label{evM}
e^{Q_l}=e^{-i\hbar\lambda_l+g(N-l)}
\ee
Let us explain that the dual system is the same because of \cite[sec.6,eq.(6.6)]{Mac}
\begin{equation}\label{deq}
{M_\lambda\Big(e^{-i\hbar\mu_l-gl};e^{-i\hbar},e^g\Big)\over M_\lambda\Big(e^{-gl};e^{-i\hbar},e^g\Big)}={M_\mu\Big(e^{-i\hbar\lambda_l-gl};e^{-i\hbar},e^g\Big)\over M_\mu\Big(e^{-gl};e^{-i\hbar},e^g\Big)}
\end{equation}
i.e. the Macdonald polynomials provide the self-dual solution to the eigenvalue problem with the Hamiltonians (\ref{eq:26}). Note that, in other terms, this equality can be associated with symmetricity of the Hopf link hyperpolynomial with respect to permuting its two components \cite{IK,AKMMhopf2}.

Indeed, the eigenvalues (\ref{evM}) are inevitably discrete like it was in our examples of section \ref{qdN2}. Then, while the wave function problem (\ref{evpM}),
\be\label{evpd}
e^{{(N-k)k\over 2}g}\sum_{
      \begin{smallmatrix}
        I \subset
        \{1,\ldots,N \}\\
        |I|=k
      \end{smallmatrix}
} \left[ \prod_{{m \in \{1,\ldots,N\} \backslash I}\atop{l \in I}}  \frac{\sinh {q_l-q_m+g\over 2}}{\sinh{q_l-q_m\over 2}} \right] M_\lambda\Big(e^{q_l-i\hbar\delta_{lj}};e^{-i\hbar},e^g\Big)=\nn\\
=e_k \Big(e^{Q_j}\Big)M_\lambda\Big(e^{q_l};e^{-i\hbar},e^g\Big)
=e_k \Big(e^{-i\hbar\lambda_j+g(N-j)}\Big)M_\lambda\Big(e^{q_l};e^{-i\hbar},e^g\Big)
\ee
is formulated at any continuous $q_l$, the dual eigenvalue problem can be written down only at discrete $Q_l$. To see the self-duality, it has to be considered also at discrete $q_l$, and we parameterize it as in (\ref{evM}): $q_l=-i\hbar\mu_l-gl$ with some partition $\mu$.
Then, one can note, using (\ref{deq}) and the fact that $e^{-i\hbar\p_{Q_l}}$ acts as $\lambda_l\to\lambda_l+1$ because of equation (\ref{evM}), that
$$
\hat H^D_kM_{\lambda}\Big(e^{-i\hbar\mu_l-gl};e^{-i\hbar},e^g\Big)=
$$
\be
&=&e^{{(N-k)k\over 2}g}\sum_{
      \begin{smallmatrix}
        I \subset
        \{1,\ldots,N \}\\
        |I|=k
      \end{smallmatrix}
}\left[ \prod_{{m \in \{1,\ldots,N\} \backslash I}\atop{l \in I}}  \frac{\sinh {Q_l-Q_m+g\over 2}}{\sinh{Q_l-Q_m\over 2}} \right] M_{\{\lambda_l+\delta_{lj}\}}\Big(e^{-i\hbar\mu_l-gl};e^{-i\hbar},e^g\Big)\sim\nn\\
&\stackrel{(\ref{deq})}{\sim}& e^{{(N-k)k\over 2}g}\sum_{
      \begin{smallmatrix}
        I \subset
        \{1,\ldots,N \}\\
        |I|=k
      \end{smallmatrix}
}\left[ \prod_{{m \in \{1,\ldots,N\} \backslash I}\atop{l \in I}}  \frac{\sinh {Q_l-Q_m+g\over 2}}{\sinh{Q_l-Q_m\over 2}} \right] M_{\mu}\Big(e^{-i\hbar\lambda_l-i\hbar\delta_{lj}-gl};e^{-i\hbar},e^g\Big)\sim\nn\\
&\stackrel{(\ref{evM})}{\sim}& e^{{(N-k)k\over 2}g}\sum_{
      \begin{smallmatrix}
        I \subset
        \{1,\ldots,N \}\\
        |I|=k
      \end{smallmatrix}
}\left[ \prod_{{m \in \{1,\ldots,N\} \backslash I}\atop{l \in I}}  \frac{\sinh {Q_l-Q_m+g\over 2}}{\sinh{Q_l-Q_m\over 2}} \right] M_{\mu}\Big(e^{Q_l-i\hbar\delta_{lj}};e^{-i\hbar},e^g\Big)\sim\nn\\
&{\stackrel{(\ref{evpd})}{\sim}}&e_k \Big(e^{-i\hbar\mu_j+g(N-j)}\Big)M_\mu\Big(e^{Q_l};e^{-i\hbar},e^g\Big)
\stackrel{(\ref{evM})}{=}e_k \Big(e^{-i\hbar\mu_j+g(N-j)}\Big)M_\mu\Big(e^{-i\hbar\lambda_l+g(N-l)};e^{-i\hbar},e^g\Big)\sim\nn\\
&\stackrel{(\ref{deq})}{\sim}& e_k \Big(e^{q_j}\Big)M_\lambda\Big(e^{-i\hbar\mu-gl};e^{-i\hbar},e^g\Big)
=e_k \Big(e^{q_j}\Big)M_\lambda\Big(e^{q_l};e^{-i\hbar},e^g\Big)\nn\\
\ee
Hence, the system is self-dual, as
expected.

One can demonstrate how the mysterious self-duality formula (\ref{deq}) emerges in the simplest case of $N=2$. The Macdonald polynomial in this case is described by the formula similar to (\ref{Jack}) with the numbers replaced with the quantum numbers, and non-zero is only the polynomial labeled by no more than two-line diagrams\footnote{In fact, up to a trivial $U(1)$-factor $(x_1x_2)^{\lambda_2}$, it is reduced to the Macdonald polynomial labeled by one-line Young diagrams with $\lambda=\lambda_1-\lambda_2$, the formula being much similar to (\ref{Jack}):
\be
M_{[\lambda]}(x_1,x_2;e^{-i\hbar},e^g)=\sum_{k=0}^{\lambda}x_1^{\lambda-k}x_2^{k}\prod_{j=0}^{k-1}c_{\lambda,j}(e^{-i\hbar},e^g)=
(x_1x_2)^{\lambda/2}\sum_{k=0}^{\lambda}\xi^{\lambda-2k}\prod_{j=0}^{k-1}c_{\lambda,j}(e^{-i\hbar},e^g),\ \ \ \ \ \ \ \ \ \ \xi:=\sqrt{x_1\over
x_2}\nn
\ee
}:
\be\label{M2}
M_{[\lambda_1,\lambda_2]}(x_1,x_2;e^{-i\hbar},e^g)=\sum_{k=0}^{\lambda_1-\lambda_2}
x_1^{\lambda_1-k}x_2^{\lambda_2+k}\prod_{j=0}^{k-1}c_{\lambda_1-\lambda_2,j}(e^{-i\hbar},e^g)
\nn\\
c_{\lambda,j}(u,v)={(u^{\lambda-j}-1)(u^jv-1)\over(u^{j+1}-1)
(u^{\lambda-j-1}v-1)}
\ee
The self-duality formula (\ref{deq}) in this case implies that
\be
M_\mu\Big(e^{-gl};e^{-i\hbar},e^g\Big)M_\lambda\Big(e^{-i\hbar\mu_l-gl};e^{-i\hbar},e^g\Big)
=\sum_{m=0}^{\mu_1-\mu_2}
e^{-g(\mu_1-m)}e^{-2g(\mu_2+m)}\prod_{l=0}^{m-1}c_{\mu_1-\mu_2,l}(e^{-i\hbar},e^g)\times\nn\\
\times\sum_{k=0}^{\lambda_1-\lambda_2}
e^{(-i\hbar\mu_1-g)(\lambda_1-k)}e^{(-i\hbar\mu_2-2g)(\lambda_2+k)}\prod_{j=0}^{k-1}c_{\lambda_1-\lambda_2,j}(e^{-i\hbar},e^g)=
e^{-g(\mu_1+\lambda_1)-2g(\mu_2+\lambda_2)-i\hbar (\mu_1\lambda_1+\mu_2\lambda_2)}\times\nn\\
\times\left(
\sum_{m=0}^{\mu_1-\mu_2}
e^{-gm}\prod_{l=0}^{m-1}c_{\mu_1-\mu_2,l}(e^{-i\hbar},e^g)\right)
\left(\sum_{k=0}^{\lambda_1-\lambda_2}
e^{-gk}e^{i\hbar k(\mu_1-\mu_2)}\prod_{j=0}^{k-1}c_{\lambda_1-\lambda_2,j}(e^{-i\hbar},e^g)\right)
\ee
is symmetric with respect to interchanging $(\lambda_1,\lambda_2)$ and $(\mu_1,\mu_2)$. In other words, it means that the expression
\be\label{N2sd}
\left(
\sum_{m=0}^{\mu}\prod_{l=0}^{m-1}v^{-1}c_{\mu,l}(u,v)\right)
\left(\sum_{k=0}^{\lambda}\prod_{j=0}^{k-1}v^{-1}u^{-\mu}c_{\lambda,j}(u,v)\right)
\ee
is symmetric with respect to interchanging non-negative integers $\lambda$ and $\mu$. We discuss some properties of (\ref{N2sd}) that are expected to be useful for extension to the elliptic case in the Appendix.

Our self-duality relation is for the eigenfunction at discrete points in the $x$-space,
which are efficiently dual to the discrete labeling of Young diagrams.
In fact, self-duality can be straightforwardly lifted to the level of continuous variables.
However, to this end,
one has to substitute  Macdonald polynomials by a more general symmetric function:
the Noumi-Shiraishi \cite{NS} mother function \cite{AKMM}.
It is a power series, which reduces to Macdonald {\it polynomials}
at integer values of eigenvalue parameters, which generalize the labeling by Young diagrams. That is, in the domain
$e^{q_1}\gg e^{q_2}\gg\ldots\gg e^{q_N}$,
the unique solution to the self-duality equations understood as a formal series in $\{e^{q_{i+1}-q_i}\}$, $i=1,\ldots,N-1$ is given (with generically non-integer $\lambda_i$ in (\ref{evM})) as:
\be\label{NSh}
\Psi_{Q}(q_l)=e^{\sum_i\lambda_iq_i}\cdot\prod_{1\le i<j\le N}{(e^{(i-j)g}e^{-i\hbar(\lambda_j-\lambda_i+1)};e^{-i\hbar})_\infty
\over(e^{(i-j-1)g}e^{-i\hbar(\lambda_j-\lambda_i+1)};e^{-i\hbar})_\infty}
\cdot\sum_{m_{ij}}{\cal C}_N(m_{ij},\lambda|e^{-i\hbar},e^g)e^{\sum_{i<j}^N(q_i-q_i)^{m_{ij}}}
\ee
where the sum runs over $m_{ij}$ such that $m_{ij}=0$ for $i\ge j$, $m_{ij}\in \mathbb{Z}_{\ge 0}$, and the coefficients ${\cal C}_N(m_{ij},s|q,t)$ are expressed through the Pochhammer symbols (\ref{Poch}) as
\be
\begin{array}{lc}
{\cal C}_N(m_{ij},\lambda|u,v):=&
\prod_{k=2}^N\prod_{1\le i<j\le k}{
\displaystyle{\Big(u^{\lambda_j-\lambda_i+\sum_{a>k}(m_{ia}-m_{ja})}v^{i-j+1};u\Big)_{m_{ik}}}
\over \displaystyle{\Big(u^{\lambda_j-\lambda_i+\sum_{a>k}(m_{ia}-m_{ja})}uv^{i-j};u\Big)_{m_{ik}}}}\times\nn\\
&\times\prod_{k=2}^N\prod_{1\le i\le j<k}{\displaystyle{
\Big(u^{\lambda_j-\lambda_i-m_{jk}+\sum_{a>k}(m_{ia}-m_{ja})}uv^{i-j-1};u\Big)_{m_{ik}}}
\over \displaystyle{\Big(u^{\lambda_j-\lambda_i-m_{jk}+\sum_{a>k}(m_{ia}-m_{ja})}v^{i-j};u\Big)_{m_{ik}}}}
\end{array}
\label{c}
\ee
At all $\lambda_i$ integer, the infinite sum becomes finite, and $\Psi_{Q}(q_l)$, a symmetric (Macdonald) polynomial.

\subsection{Dual elliptic RS model\label{sE}}

Hamiltonians of the $N$-body elliptic RS model are\footnote{$\theta$-function here is the standard odd $\theta$-function $\theta_1(x,\tau)$ with a rescaling of the variable: $\theta(2\pi ix)=\theta_1(x,\tau)$, $\mathfrak{q}=e^{2\pi i\tau}$. This rescaling provides more convenient limit to the trigonometric RS case.}
\begin{equation}
  \label{eR}
   \hat  H_k^{eR}=e^{{(N-k)k\over 2}g}\sum_{
      \begin{smallmatrix}
        I \subset
        \{1,\ldots,N \}\\
        |I|=k
      \end{smallmatrix}
} \prod_{l \in I}\left[ \prod_{m \in \{1,\ldots,N\} \backslash I}  \frac{\theta \Big({q_l-q_m+g\over 2}\Big)}{\theta\Big({q_l-q_m\over 2}\Big)} \right] \prod_{j \in I} e^{-i\hbar \p_{q_j}}
\end{equation}
Their duals are unknown. However, the wave functions of the dual Hamiltonians are known by duality \cite{MMZ}: one can construct a set of functions labeled by partitions $\psi_\lambda(q_i)$ such that they are wave functions of the elliptic RS Hamiltonians (\ref{eR}) {\bf acting on partitions (parameterizing the eigenvalues of the dual system)}. In other words,
\be
\hat  H_k^{eR}(Q_l)\psi_{\lambda(Q_l)}(q_j)=e_k(q_l)\psi_{\lambda(Q_l)}(q_j)
\ee
These wave functions explicitly are
\be
\psi_{\lambda(Q_l)}(q_j)\Big|_{Q_l = q^{\lambda_l}t^{1-l}} =  \mathfrak{F}(q^{\lambda_l}t^{-l}) E_{\lambda}(e^{q_j})\nn\\
 \mathfrak{F}(x_l)= \prod_{l<j} \left[x_j^{i\frac{g}{\hbar}} \prod_{l<j\le N}{\Gamma\Big(e^{q_l-q_j};e^{-i\hbar},\mathfrak{q}\Big)_\infty\over
\Gamma\Big(e^{q_l-q_j+g};e^{-i\hbar},\mathfrak{q}\Big)_\infty } \right]
\ee
where $\Gamma(x;z,w)=\prod_{n,m \geq 0}{\left(1 - x z^n w^m \right)\over \left(1 - x^{-1} z^{n+1} w^{m+1}
    \right)}$ is the elliptic Gamma-function (\ref{EG}), and $E_\lambda(x_l)$ are symmetric polynomials of $x_l$ manifestly described in \cite{GNS}, the first few of them in terms of power sums $\mathfrak{p}_k=\sum_jx_j^k$ are
\be
  \label{eq:1}
  E_{[1]}\{\mathfrak{p}_k\} &=& \mathfrak{p}_1\nn\\
  E_{[1,1]}\{\mathfrak{p}_k\} &=& \frac{1}{2} (\mathfrak{p}_1^2 - \mathfrak{p}_2)\nn\\
  E_{[2]}\{\mathfrak{p}_k\} &=& \frac{2-\sigma(0)}{2} \mathfrak{p}_1^2 +
  \frac{\sigma(0)}{2}\mathfrak{p}_2\nn\\
  E_{[1,1,1]}\{\mathfrak{p}_k\} &=&\frac{\mathfrak{p}_3}{3} - \frac{\mathfrak{p}_2\mathfrak{p}_1}{2} + \frac{\mathfrak{p}_1^3}{6}\nn\\
  E_{[2,1]}\{\mathfrak{p}_k\} &= &\frac{1}{6} (3 - \sigma(0) \sigma(g))
  \mathfrak{p}_1^3 + \frac{1}{2} (\sigma(0) \sigma(g) - 1) \mathfrak{p}_1 \mathfrak{p}_2 -
  \frac{1}{3}\sigma(0) \sigma(g) \mathfrak{p}_3\nn\\
  E_{[3]}\{\mathfrak{p}_k\} &=& \left( 1 - \frac{\sigma(0)}{2}  -
    \frac{\sigma(-i\hbar)}{2}  + \frac{\sigma(0)  \sigma(g) \sigma(-i\hbar)}{6}  \right)
  \mathfrak{p}_1^3 + \frac{\sigma(0) +  \sigma(-i\hbar)
  -\sigma(0) \sigma(g) \sigma(-i\hbar)}{2} \mathfrak{p}_1 \mathfrak{p}_2+\nn\\ &+& \frac{\sigma(0) \sigma(g)\sigma(-i\hbar) }{3} \mathfrak{p}_3
\ee
where
\begin{equation}
  \label{eq:2}
  \sigma(x) =\frac{\theta (-i\hbar+ x) \theta(2g+ x)}{ \theta(-i\hbar+g+ x)\theta(g+x)}
\end{equation}
In particular, at $N=2$ and $x_2=x_1^{-1}=x^{-1}$, these polynomials are
\be
E_{[n]}=\sum_{k=0}^nx^{n-2k}c_{k,n}^{ell}
\ee
and
\be
c_{1,n}^{ell}&=&C_n^1-C_{n-2}^0\sum_{a_1=0}^{n-2}\sigma(-i\hbar a_1)\nn\\
c_{2,n}^{ell}&=&C_n^2-C_{n-2}^1\sum_{a_1=0}^{n-2}\sigma(-i\hbar a_1)+C_{n-4}^0\sum_{a_1=0}^{n-2}\sum_{a_2=0}^{a_1-2}
\sigma(-i\hbar a_1)\sigma(-i\hbar a_2)\nn\\
c_{3,n}^{ell}&=&C_n^3-C_{n-2}^2\sum_{a_1=0}^{n-2}\sigma(-i\hbar a_1)+C_{n-4}^1\sum_{a_1=0}^{n-2}\sum_{a_2=0}^{a_1-2}
\sigma(-i\hbar a_1)\sigma(-i\hbar a_2)-C_{n-6}^0\sum_{a_1=0}^{n-2}\sum_{a_2=0}^{a_1-2}\sum_{a_3=0}^{a_2-2}
\sigma(-i\hbar a_1)\sigma(-i\hbar a_2)\sigma(-i\hbar a_3)\nn\\
\ldots
\ee
where $C_n^k$ are binomial coefficients.
Note that these formulas slightly remind those emerged earlier within a simpler framework of \cite{Gleb}.

\subsection{Duality and KS Hamiltonians}

In the previous subsection, we constructed the wave function of the dual elliptic RS system, but the Hamiltonians are still remained to be constructed. Here we consider auxiliary KS Hamiltonians constructed in \cite{KS}, which are not self-dual. In fact, we consider their elliptic$_p$-trigonometric$_q$ degeneration: their corresponding wave functions $P_\lambda$ are related to the polynomials $E_\lambda$ introduced above by bi-orthogonality relations. This will allow us to make a step towards constructing the Hamiltonians for the dual elliptic RS Hamiltonians in the next section.

The KS Hamiltonians are
\be\label{KS}
\hat {{H}}^{KS}_a(q_i|\mathfrak{q},\hbar,g):=\hat{\mathfrak{O}}_0^{-1}(q_i|\mathfrak{q},\hbar,g)\cdot\hat{\mathfrak{O}}_a(q_i|\mathfrak{q},\hbar,g),\ \ \ \ \ \ \ a=1\ldots N-1
\ee
where $\hat{\mathfrak{O}}_a$ is read from
\be\label{Oop}
\hat{\mathfrak{O}}(z|q_i|\mathfrak{q},\hbar,g)&=&\sum_{k\in\mathbb{Z}}\hat{\mathfrak{O}}_k(q_i|\mathfrak{q},\hbar,g)z^k:=\nn\\
&=&\sum_{k_1,\ldots,k_N\in\mathbb{Z}}
z^{\sum_j k_j}\mathfrak{q}^{\sum_l k_l(k_l-1)/2}\prod_{l<m}\sinh{g(k_l-k_m)+q_l-q_m\over 2}
\prod_{j=1}^ne^{-i\hbar k_j\p_{q_j}}\nn\\
\ee
In fact, there is a periodicity in the index $k$ of the Hamiltonian \cite{KS} so that only the first $N-1$ Hamiltonians are  independent.

In the simplest case of $N=2$,
\be\label{H2e}
\hat{\mathfrak{O}}_0=\sum_{n\in\mathbb{Z}}\mathfrak{q}^{n^2}\sinh{q_1-q_2+2gn\over 2}
e^{-i\hbar n\p_1+i\hbar n\p_2}\nn\\
\hat{\mathfrak{O}}_1=\sum_{n\in\mathbb{Z}}\mathfrak{q}^{n^2-n}\sinh{q_1-q_2+g(2n-1)\over 2}
e^{-i\hbar n\p_1+i\hbar (n-1)\p_2}
\ee
and one would better write down the eigenvalue equation in the form
\be\label{111}
\hat{\mathfrak{O}}_1P_R(e^{q_1},e^{q_2};\hbar,g)=\Lambda_R\hat{\mathfrak{O}}_0P_R(e^{q_1},e^{q_2};\hbar,g)
\ee

Generally, the eigenfunctions of the Hamiltonians (\ref{KS}) are symmetric polynomials \cite{GNS,MMZ}
\begin{align}
P_{[1]}\{\mathfrak{p}_k\} &= \mathfrak{p}_1 \nn \\
P_{[1,1]}\{\mathfrak{p}_k\} &= \frac{\mathfrak{p}_1^2-\mathfrak{p}_2}{2}
\nn\\
P_{[2]}\{\mathfrak{p}_k\} &= \frac{2-\zeta(0)}{2}\mathfrak{p}_2 + {\zeta(0)\over 2}\mathfrak{p}_1^2
\nn \\
P_{[1,1,1]}\{\mathfrak{p}_k\} &= \frac{\mathfrak{p}_3}{3} - \frac{\mathfrak{p}_2\mathfrak{p}_1}{2} + \frac{\mathfrak{p}_1^3}{6}\nn\\
P_{[2,1]}\{\mathfrak{p}_k\} &= {\zeta(0)+\zeta(g)-3\over 3}\mathfrak{p}_3-{\zeta(0)+\zeta(g)-2\over 2}\mathfrak{p}_2\mathfrak{p}_1
+{\zeta(0)+\zeta(g)\over 6}\mathfrak{p}_1^3
\nn \\
P_{[3]}\{\mathfrak{p}_k\} &= \left(1-\zeta(-i\hbar)\zeta(0) + \frac{\zeta(-i\hbar)\zeta(0)^2}{3}\right)\mathfrak{p}_3
+ \zeta(-i\hbar)\zeta(0)\left(1-\frac{\zeta(0)}{2}\right)\mathfrak{p}_2\mathfrak{p}_1 + \frac{\zeta(-i\hbar)\zeta(0)^2}{6}\mathfrak{p}_1^3
\end{align}
where
\be
\zeta(z) =   \frac{\theta(-2i\hbar +z)\theta(g+z)}{\theta(-i\hbar+g +z)\theta(-i\hbar+ z)}=\sigma(z)\Big|_{g\leftrightarrow -i\hbar}
\ee
and $\mathfrak{p}_k=\sum_ie^{kq_i}$.
They are related to $E$-polynomials (\ref{eq:1}) via
\be
E_{\lambda}^{(-i\hbar,g)}\{\mathfrak{p}_k\}:=P_{\lambda^\vee}^{(g,-i\hbar)\perp}\{(-1)^{k+1}\mathfrak{p}_k\}
\ee
where the superscript $\perp$ indicates the set of symmetric polynomials $P^\perp_\lambda$ orthogonal to $P_\mu$ with respect to the Schur scalar product \cite{Mac}, i.e. such that $\Big<S_\mu\Big|S_\lambda\Big>=\delta_{\lambda\mu}$, and the superscript $\vee$ denotes the transposed Young diagram. Conjugation of the operator with respect to this scalar product corresponds to the replace $\mathfrak{p}_k\leftrightarrow k{\p\over\p\mathfrak{p}_k}$. Hence, in order to find out the dual elliptic RS Hamiltonians, it is sufficient to rewrite Hamiltonians (\ref{KS}) in terms of variables $p_k$ and to make this replace. We will do this in the section. However, it gives rise only to dual elliptic RS Hamiltonians in term of $\mathfrak{p}_k$-variables, but not in terms of coordinates $q_i$.

\section{Ambiguities of quantum duality problem\label{amb}}

As we already discussed in sec.2, there are various ambiguities in the duality problem. At the classical level, first of all, one can choose a set of dual Hamiltonians in (\ref{duality}) in many way, since functions of Hamiltonians are still Hamiltonians. Second,  there is a freedom in the anti-canonical transformation in (\ref{dualH}). At the same time, when dealing with the self-dual system, this freedom disappears.

At the quantum level, the second ambiguity is hidden in choosing the proper wave function. However, there emerges yet another ambiguity as compared with the classical system: there is a freedom in ordering of the operators in quantum Hamiltonians. To understand the persisting problems in finding the DELL system(s), what we want to emphasize is that the definition of quantum duality may depend on this additional choice, and, perhaps, in a non-naive way.

To understand this point better, let us discuss the choice of wave functions in the quantum (self-)duality problem. First of all, let us note that selection rules for wave functions come by a standard procedure in quantum mechanics often hidden under a specification of boundary conditions. In general the differential operator of order $n$ has $n$-dimensional space of eigenfunctions.
For instance, the trigonometric Calogero Hamiltonian (\ref{qCS}) has a general solution which is a linear combination of Legendre functions of two kinds, ${\bf P}_a^b(z)$ and ${\bf Q}_a^b(z)$:
\be
\psi_Q(q)=C_1\cdot\sqrt{\sinh q}\cdot {\bf P}_{-{1\over 2}+iQ}^{-{1\over 2}+\beta}(\cosh q)+
C_2\cdot\sqrt{\sinh q}\cdot {\bf Q}_{-{1\over 2}+iQ}^{-{1\over 2}+\beta}(\cosh q)
\ee
and only their particular linear combination at the {\it discrete} set of values of $Q$ gives rise to the Jack {\it polynomial} (\ref{Jack})
that we discussed in sec.4. Lifting duality to the level of generic Legendre functions even in this case is a separate story.

Things become even more involved when we go to difference equations, i.e to RS systems:
the ambiguity increases, and there is a freedom up to $N$ arbitrary periodic functions.
As we explained in sec.6.1, the trigonometric self-duality of the Macdonald polynomials is only a part of the story: in this case, we solve equation (\ref{evpM}),
\be\label{bs1}
\hat  H_k^q\Psi_{Q}(q_l)=e_k \Big(e^{Q_j}\Big)\Psi_{Q}(q_l)
\ee
where the superscript $q$ of the Hamiltonian indicates that it acts on variables $q_i$, and realize that its solutions are the Macdonald polynomials (\ref{efM}) also at the {\it discrete} set of values of $Q$. However, this case is self-dual, hence, we understand how to choose the self-dual solution at arbitrary values of $Q$: we impose the second equation similar to (\ref{evpM}) but with the Hamiltonian acting on the variables $Q_i$ instead of $q_i$:
\be\label{bs2}
\hat  H_k^Q\Psi_{Q}(q_l)=e_k \Big(e^{q_j}\Big)\Psi_{Q}(q_l)
\ee
This pair of equations (\ref{bs1}), (\ref{bs2}) has a unique solution (\ref{NSh}) and, at the first glance, provides a unique solution to the self-duality problem. However, in this case, amusingly or not, there exists at least one another set of self-dual RS Hamiltonians giving rise to a very different type of eigenfunctions: multiple integrals of the Mellin-Barnes type \cite{Kh0,Kh,Kh2,Kh3}.

The point is that, as we emphasized above, there is a problem of ordering of quantum Hamiltonians, and the Hamiltonians (\ref{eq:26}) is only one of many possible choices of quantum Hamiltonians for the same classical $N$-particle Ruijsenaars system. One can consider other quantum Hamiltonians and, accordingly, construct a different set of self-dual eigenfunctions. In \cite{Kh0,Kh,Kh2,Kh3}, the authors constructed eigenfunctions of the Hamiltonians
\be\label{HRS2}
   \hat  {\widetilde H}_k=e^{{(N-k)k\over 2}g}\sum_{
      \begin{smallmatrix}
        I \subset
        \{1,\ldots,N \}\\
        |I|=k
      \end{smallmatrix}
} \prod_{{m \in \{1,\ldots,N\} \backslash I}\atop{l \in I}}  \sqrt{ \frac{\sinh {q_l-q_m+g\over 2}}{\sinh{q_l-q_m\over 2}}}
\cdot\prod_{j \in I} e^{-i\hbar \p_{q_j}}\cdot
\prod_{{m \in \{1,\ldots,N\} \backslash I}\atop{l \in I}}  \sqrt{ \frac{\sinh {q_l-q_m+g\over 2}}{\sinh{q_l-q_m\over 2}}}
\ee
with pure imaginary $g$. Then, from formulas \cite[Eqs.(1.7),(1.48),(1.50)]{Kh}, it follows that the function
\be
\sqrt{\mu(q,g)\mu(\lambda,-i\hbar-g)}\Psi_{\lambda}^{R}(q)
\ee
solves the self-duality problem for the Hamiltonians (\ref{HRS2}). Here we used the notation of \cite{Kh}:
$\mu(q,g)$ is the Macdonald orthogonality measure (a deformed square of the Vandermonde determinant)\footnote{It requires a regularization in the case of $g$ pure imaginary considered in \cite{Kh0,Kh,Kh2,Kh3}. The correspondence of notations is as follows: $q_i={2\pi\hat q_i\over\omega_2}$, $\hbar={2\pi\omega_1\over\omega_2}$, $g=-{2\pi i\hat g\over\omega_2}$, where the hatted letters denote the quantities from the papers by N. Belousov et al.} \cite[Eq.(9.2)]{Mac},
$\Delta\left(e^{q_i};e^{-i\hbar},e^{g}\right)=\mu(q,g)$,
and the functions $\Psi_{\lambda}^{R}(x)$ are given explicitly as multiple integrals of
Mellin-Barnes type of the combinations of double signs functions, \cite[Eq.(1.23)]{Kh2} (see also \cite{HRui}).

However, solutions in this case are looking not that immediately related to symmetric polynomials.

One probably should not expect that these two cases exhaust the full set of possibilities in this particular case:
it can easily happen that there is a self-dual solution to arbitrary choice of ordering in the quantization of classical
self-dual system. The problem deserves attention, because for double elliptic generalizations
there are no chances of selecting a ``nice" eigenfunction systems by choosing {\it polynomials}:
there are no polynomial eigenfunctions for the Hamiltonian elliptic in coordinates.
Thus understanding of the DELL system requires better insight in this kind of ambiguities.

\section{Hamiltonians in terms of $\mathfrak{p}_k$-variables\label{spk}}

In the previous section, we could see that it was sometimes much simpler to express the wave functions that turned out to be symmetric polynomials in terms of power sums $\mathfrak{p}_k$. If the number of degrees of freedom (the number of $q_i$) is large: $N\to\infty$, one can consider these power sums as good variables instead of symmetric variables (see \cite{V} for an accurate description of this procedure). Hence, it is useful to rewrite the Hamiltonians in these terms. One could try to construct a reformulation of the duality map in terms of the $\mathfrak{p}_k$-variables. Unfortunately, this is not possible: even if the wave function could be a symmetric polynomial, its dual is typically not a polynomial, and the formulation in terms of $\mathfrak{p}_k$-variables is unavailable. Still, it is useful to construct Hamiltonians in these terms in those cases when polynomial wave functions do exist. This is what we discuss in this section.

\subsection{A warm-up example: Schur polynomials}

The Schur polynomials give rise to a set of wave functions of trivial free Hamiltonians: the Hamiltonians of the trigonometric CS model at the free fermion point. They are called generalized cut-and-join operators and can be manifestly described in the following way \cite{MMN}: one introduces a matrix $H$ such that $\mathfrak{p}_k=\Tr H^k$ (Miwa variables) and constructs the operators\footnote{Hereafter, by the matrix derivative, we imply the derivative w.r.t. matrix elements of the transposed matrix: $\left(\frac{\partial}{\partial H}\right)_{ij}=\frac{\partial}{\partial H_{ji}}$.}
\be
\hat W_\Delta := \ :\prod_{a=1}^{l_\Delta}  \tr \left(H\frac{\p}{\p H_{tr}}\right)^{\Delta_a}:
\ee
where the normal ordering $:\ldots :$ implies all the derivatives put to the right. Then, the Schur functions $S_\lambda\{\mathfrak{p}_k\}$ are the eigenfunctions of these operators,
\be
\hat W_\Delta\ S_\lambda\{\mathfrak{p}_k=\Tr H^k\}=\phi_\lambda(\Delta)\ S_\lambda\{\mathfrak{p}_k=\Tr H^k\}
\ee
and the eigenvalues are
\be
\phi_\lambda(\Delta)=\sum_{\mu\vdash |\Delta|}{S_{\lambda/\mu}\{\delta_{1,k}\}\over S_{\lambda}\{\delta_{1,k}\}}\
{Ch_\mu(\Delta)\over z_\Delta}
\ee
Here $Ch_\mu(\Delta)$ is the value of character of the symmetric group $S_{|\Delta|}$ in representation $\mu$ on the conjugacy class labeled by the Young diagram $\Delta$, $z_\Delta$ is the standard symmetric factor of the Young diagram (order of the automorphism) \cite{Fulton}, and the sum runs over partitions of the size $|\Delta|$.

The simplest non-trivial example of the Hamiltonian is
\be
\hat W_{[2]}={1\over 2}\sum_{a,b>0}\left((a+b)\mathfrak{p}_a\mathfrak{p}_a{\p\over\p\mathfrak{p}_{a+b}}+
ab\mathfrak{p}_{a+b}{\p\over\p\mathfrak{p}_a\p\mathfrak{p}_b}\right)
\ee
It can be rewritten in terms of eigenvalues $h_i$ of the matrix $H$ instead of $\mathfrak{p}_k$-variables, the result reads
\be
\hat W_{[2]}={1\over 2}\sum_ih_i^2{\p^2\over\p h_i^2}+{1\over 2}\sum_{i\ne j}{h_ih_j\over h_i-h_j}\left({\p\over\p h_i}-{\p\over\p h_j}\right)
\ee
In fact, it is better to consider the Hamiltonian $\hat H_2=\hat W_{[2]}+{2N-1\over 2}\ \hat W_{[1]}$ so that
\be\label{Van}
\Delta(h)\cdot\hat H_2\cdot\Delta(h)^{-1}={1\over 2}\sum_ih_i^2{\p^2\over\p h_i^2}+{1\over 2}\sum_ih_i{\p\over\p h_i}-{N(N-1)(2N-1)\over 12}=
{1\over 2}\sum_i{\p^2\over\p q_i^2}-{N(N-1)(2N-1)\over 12}\nn\\
\ee
where $\Delta(x)=\prod_{i<j}(x_i-x_j)$ is the Vandermonde determinant, and $h_i=e^{q_i}$,
i.e. the polynomial $\Delta(x) S_\lambda(x_i)$ is a wave function of the free system Hamiltonian (hence, the name free fermion point).

\subsection{Trigonometric CS model: Jack polynomials}

Similarly, there is a set of commuting Hamiltonians in terms of variables $\mathfrak{p}_k$ in the interacting trigonometric CS model, and as we observed in sec.\ref{qdN2} their eigenfunctions are the Jack polynomials. These Hamiltonians can be no longer realized in terms of matrices, but can be constructed again in terms of ``eigenvalues", which are exponentials of the coordinates, $h_i=e^{q_i}$ and in terms of $\mathfrak{p}_k$-variables.

In terms of coordinates, the simplest Hamiltonian is
\be\label{qCS2}
\hat W_{[2]}^\beta={1\over 2}\sum_ih_i^2{\p^2\over\p h_i^2}+{\beta\over 2}\sum_{i\ne j}{h_ih_j\over h_i-h_j}\left({\p\over\p h_i}-{\p\over\p h_j}\right)
\ee
The eigenfunctions of this Hamiltonian are the Jack polynomials $J_\lambda(h_i)$, the eigenfunctions being
\be
\hat W_{[2]}^\beta\cdot J_\lambda(h_i)&=&\Lambda_\lambda\cdot J_\lambda(h_i)\\
\Lambda_\lambda&=&{1\over 2}\sum_m\left[\Big(\lambda_m-\beta m+\beta-{1\over 2}\Big)^2-\Big(-\beta m+\beta-{1\over 2}\Big)^2\right]={1\over 2}\sum_m\lambda_m\Big(\lambda_m-2\beta m+2\beta -1\Big)\nn
\ee
In order to generate the standard form of the trigonometric CS Hamiltonian (\ref{qCS}), one again has to multiply the Jack polynomials with the Vandermonde determinant and to add again $\hat W_{[1]}$:
\be
\hat H_2^\beta=\hat W_{[2]}^\beta+\left(N\beta-\beta+{1\over 2}\right)\ \hat W_{[1]}\nn
\ee
\be
\Delta(h)^\beta\cdot\hat H_2^\beta\cdot\Delta(h)^{-\beta}&=&
{1\over 2}\sum_ih_i^2{\p^2\over\p h_i^2}+{1\over 2}\sum_ih_i{\p\over\p h_i}
-{\beta(\beta-1)\over 2}\sum_{i\ne j}{h_ih_j\over(h_i-h_j)^2}-{\beta^2 N(N-1)(2N-1)\over 12}=\nn\\
&\stackrel{h_i=e^{q_i}}{=}&
{1\over 2}\sum_i{\p^2\over\p q_i^2}-{\beta(\beta-1)\over 2}\sum_{i\ne j}{1\over\sinh^2{q_i-q_j\over 2}}
-{\beta N(N-1)(2N-1)\over 12}
\ee
i.e. the polynomial $\Delta(h)^\beta J_\lambda(e^{q_i})$ is a wave function of the trigonometric CS Hamiltonian.
The multiplication with the Vandermonde determinant corresponds to another choice of the variables $P_i$ (the freedom in the definition of the duality map).

In terms of $\mathfrak{p}_k$-variables, the Hamiltonian (\ref{qCS2}) is
\be
\hat W_{[2]}^\beta={1\over 2}\sum_{a,b>0}\left(\beta(a+b)\mathfrak{p}_a\mathfrak{p}_a{\p\over\p\mathfrak{p}_{a+b}}+
ab\mathfrak{p}_{a+b}{\p\over\p\mathfrak{p}_a\p\mathfrak{p}_b}\right)+{1-\beta\over 2}\sum_k (k-1)k\mathfrak{p}_k \dfrac{\partial}{\partial \mathfrak{p}_k}
\ee

\subsection{Trigonometric RS model: Macdonald polynomials}

Similarly to the trigonometric CS system, one can realize Hamiltonians of the trigonometric RS system both in terms of coordinates, (\ref{eq:26}) and in terms of $\mathfrak{p}_k$-variables. The simplest Hamiltonian is
\be\label{Hq}
   \hat  H_1=e^{{N-1\over 2}g}\sum_{l=1}^N \left[
   \prod_{m\ne l}  \frac{\sinh {q_l-q_m+g\over 2}}{\sinh{q_l-q_m\over 2}} \right] e^{-i\hbar \p_{q_l}}
\ee
in terms of coordinates, and\footnote{Hereafter, we normalize the contour integral in such a way that $\oint_0{dz/z}=1$.}
\be\label{HpRS}
\hat {\cal H}_1=\oint_0{dz\over z}
\exp\left(\sum_{k>0}{(1-e^{-kg})\mathfrak{p}_kz^k\over k}\right)\cdot\exp\left(\sum_{k>0}{e^{-i\hbar k}-1\over z^k}
{\partial\over\partial \mathfrak{p}_k}\right)
\ee
in terms of $\mathfrak{p}_k$-variables. The eigenfunctions of these Hamiltonians are the Macdonald polynomials $M_\lambda$,
\be
\hat {\cal H}_1\cdot M_\lambda(q_l)=\Lambda_1\cdot M_\lambda(q_l)\nn\\
\hat {H}_1\cdot M_\lambda\{\mathfrak{p}_k\}=\Lambda\cdot M_\lambda\{\mathfrak{p}_k\}
\ee
and the eigenvalues are (see (\ref{evM}))
\be
\Lambda_1=e^{g(N-1)}\sum_l{e^{-i\hbar\lambda_l}\over e^{(l-1)g}}
\ee
and
\be
\Lambda=1+2\sinh {g\over 2}\sum_l{e^{-i\hbar\lambda_l}-1\over e^{(l-1)g}}=
1+2\sinh {g\over 2}\sum_l{e^{-i\hbar\lambda_l}\over e^{(l-1)g}}-e^{g\over 2}
\ee
Comparing these eigenvalues, one observes that the Hamiltonians are related by a simple linear transformation.
This is because they both are the lowest Hamiltonians.

A simple choice of higher Hamiltonians in terms of coordinates is given by (\ref{evpM}). There are many natural ways to construct higher Hamiltonians in terms of $\mathfrak{p}_k$-variables, some of them described in \cite[Sec.2.2]{MMgenM}. Note that the eigenfunctions of these Hamiltonians are just the Macdonald polynomials, while the eigenfunctions of the trigonometric CS Hamiltonians are the Jack polynomials multiplied with the Vandermonde determinant. This means that, in order to go to the trigonometric CS model (\ref{qCS}), one has not only to perform the limit $t=q^\beta$, $q\to 1$, but also to ``rotate" the Hamiltonians with the Vandermonde factor as in (\ref{Van}).
Note also that, upon choosing $t=q$, one obtains the Hamiltonians that have the Schur functions as their eigenfunctions.

\subsection{Dual elliptic RS model: $E_\lambda$-polynomials\label{El}}

Consider now the case of $N=2$ of the KS Hamiltonians (\ref{H2e}). Following \cite{AK,Zen}, one can rewrite (\ref{111})
in the form
\be
\hat {\mathfrak{H}}_1P_\lambda\{\mathfrak{p}_k;\hbar,g\}
=\Lambda_\lambda\hat {\mathfrak{H}}_0P_\lambda\{\mathfrak{p}_k;\hbar,g\}
\ee
at $p_k=e^{kq_1}+e^{kq_2}$,
where the Hamiltonians (they are related with those in (\ref{H2e}) by a simple rescaling) has the integral form
\be
\hat {\mathfrak{H}}_0=\sum_{n\in\mathbb{Z}}{\Big(w^ne^{5g}\Big)^{n}\over\sinh^2 ng}
\oint{dz_1\over z_1}{dz_2\over z_2}{z_1-z_2\over z_1-e^{2ng}z_2}
\exp\left(\sum_k{1-e^{-2nkg}\over k}(z_1^{-k}+z_2^{-k})p_k\right)\times\nn\\
\times\exp\left(\sum_k\Big[(e^{-i\hbar nk}-1)z_1^{k}+(e^{i\hbar nk}-1)z_2^{k}\Big]\p_{p_k}\right)\nn\\
\hat {\mathfrak{H}}_1=e^{2g}\sum_{n\in\mathbb{Z}}{\Big(w^ne^{5g}\Big)^{n-1}\over\sinh^2 (n-1/2)g}
\oint{dz_1\over z_1}{dz_2\over z_2}{z_1-z_2\over z_1-e^{(2n-1)g}z_2}
\exp\left(\sum_k{1-e^{-(2n-1)kg}\over k}(z_1^{-k}+z_2^{-k})p_k\right)\times\nn\\
\times\exp\left(\sum_k\Big[(e^{-i\hbar nk}-1)z_1^{k}+(e^{i\hbar (n-1)k}-1)z_2^{k}\Big]\p_{p_k}\right)
\ee
where the both $z_i$ run over the same integration contour surrounding the both points $e^{q_1}$ and $e^{q_2}$. Each of the two integrals can be deformed to the integrals around points $0$, $\infty$ and $z_1=e^{2ng}z_2$ for the summand of $\hat {\mathfrak{H}}_0$, and $z_1=e^{(2n-1)g}z_2$ for the summand of $\hat {\mathfrak{H}}_1$.
Extension to $N>2$ and to higher Hamiltonians is immediate, however, the number of integrations increases along with increasing $N$.

Another way to obtain an explicit expression for the KS Hamiltonians in terms of times suitable at any $N$ is as follows: a generating function of the KS Hamiltonians (\ref{Oop}) can be rewritten in a determinant form~\cite{GZ}. For the sake of convenience, we denote $\hat{\mathfrak{O}}(e^{u}|q_i|\mathfrak{q},\hbar,g)$ in (\ref{Oop}) through $\hat{\mathcal{O}}(u)$ and use slightly different normalization so that, in the ell$_p$-trig$_q$ case, the generating operator has the determinant representation
\begin{equation}
  \label{eq:8}
\hat{\mathcal{O}}(u) = \frac{1}{\Delta(e^q)}\det_{1\leq k,l
    \leq N} \left[ e^{(N-k)q_l}
 { \theta_{\tau}\Big(u +(1-k)g-i\hbar\partial_{q_l}\Big)\over  \theta_{\tau}\Big(u +(1-k)g\Big)} \right]
\end{equation}
where $u$ is a generating parameter, and, for the sake of brevity, we defined $\theta_\tau(x)=\prod_{n=0}\Big(1-e^{2\pi i(n\tau+x)}\Big)\Big(1-e^{2\pi i((n+1)\tau-x)}\Big)\sim e^{\pi ix}\theta_1(x,\tau)$.
This operator (\ref{eq:8}) acts as a triangular matrix in the basis of monomial symmetric
polynomials $m_{\lambda}(e^{q_i})$, moreover, this matrix has a block-diagonal form with blocks at each level $|\lambda|$. However, since $\hat{\mathcal{O}}(u)$ for different values
of $u$ \emph{do not commute,} the eigenfunctions in general
\emph{depend on $u$.} In order to get commuting Hamiltonians, one should take the ratio of the
generating operators $\hat{\mathcal{O}}(u)$ at two
different values of $u$:
\begin{equation}
  \label{eq:18}
  \hat H(v,u) = \hat {\mathcal{O}}(v) \Big(\hat {\mathcal{O}}(u)\Big)^{-1}.
\end{equation}

For instance, the matrix of the operators $\hat{\mathcal{O}}(v)$ and $\hat H(v,u)$
in the basis of $m_{\lambda}$ at the first and second levels at $N=2$ read
\be
  \label{eq:19'}
\hat{\mathcal{O}}(v) \left(
  \begin{array}{c}
  m_{[1]}\\
    m_{[1,1]}\\
    m_{[2]}
  \end{array}
 \right)
 =\left(
    \begin{array}{ccc}
    {\theta_{\tau}(v-i\hbar)\over\theta_{\tau}(v)} &0&0\\
    0& {\theta_{\tau}(v-i\hbar)\theta_{\tau}(v-i\hbar-g)\over\theta_\tau(v)\theta_\tau(v-g)}  &0\\
     0&  {\theta_{\tau}(v-2i\hbar )\over\theta_{\tau}(v)} -  {\theta_{\tau}(v-2i\hbar-g)\over\theta_\tau(v-g)} &  {\theta_{\tau}(v-2i\hbar )\over\theta_\tau(v)}
    \end{array}
\right) \left(
  \begin{array}{c}
  m_{[1]}\\
    m_{[1,1]}\\
    m_{[2]}
  \end{array}
 \right)\nn
\ee
and
\begin{equation}
  \label{eq:19}
\hat H(v,u) \left(
  \begin{array}{c}
  m_{[1]}\\
    m_{[1,1]}\\
    m_{[2]}
  \end{array}
 \right)=
  \left(
    \begin{array}{ccc}
   {\eta_{1,0}(v)\over\eta_{1,0}(u)}&0&0\\
      0& {\eta_{1,1}(v)\over\eta_{1,1}(u)} &0\\ 0&
     {\eta_{2,0}(v)\eta_{-1,1}(qu)\over\eta_{2,0}(u)} -{\eta_{0,2}(v)\over\eta_{1,1}(u)}
  &  \ \ \ {\eta_{2,0}(v)\over\eta_{2,0}(u)}
    \end{array}
\right) \left(
  \begin{array}{c}
  m_{[1]}\\
    m_{[1,1]}\\
    m_{[2]}
  \end{array}
 \right)
\end{equation}
where we denoted
\be
\eta_{m,n}(x):={\theta_\tau(x-i\hbar m)\theta_\tau(x-i\hbar n-g)
\over \theta_\tau(x)\theta_\tau(x-g)}=\ \ \ \ \ \ \ \ \\
=\prod_{n=0}{\Big(1-e^{2\pi i(n\tau+x-i\hbar m)}\Big)\Big(1-e^{2\pi i((n+1)\tau-x+i\hbar m)}\Big)\Big(1-e^{2\pi i(n\tau+x-i\hbar n-g)}\Big)\Big(1-e^{2\pi i((n+1)\tau-x+i\hbar n+g)}\Big)\over
\Big(1-e^{2\pi i(n\tau+x)}\Big)\Big(1-e^{2\pi i((n+1)\tau-x)}\Big)\Big(1-e^{2\pi i(n\tau+x-g)}\Big)\Big(1-e^{2\pi i((n+1)\tau-x+g)}\Big)}\nn
\ee

Since
\be
 m_{[1,1]}={\mathfrak{p}_1^2\over 2}-{\mathfrak{p}_2\over 2},\ \ \ \ \ \ \ \ \
    m_{[2]}=\mathfrak{p}_2
\ee
one immediately obtains from (\ref{eq:19}) that
\be\label{Hp}
\hat H(v,u)&=&{\eta_{1,0}(v)\over\eta_{1,0}(u)}\cdot\mathfrak{p}_1{\p\over\p\mathfrak{p}_1}+\nn\\
&+&{1\over 4}\left({\eta_{2,0}(v)\eta_{-1,1}(qu)\over\eta_{2,0}(u)}+{2\eta_{1,1}(v)-\eta_{0,2}(v)\over\eta_{1,1}(u)}
-{\eta_{1,0}(v)\over\eta_{1,0}(u)}\right)
\cdot\mathfrak{p}_1^2{\p^2\over\p\mathfrak{p}_1^2}-\nn\\
&-&{1\over 4}\left({\eta_{2,0}(v)\eta_{-1,1}(qu)\over\eta_{2,0}(u)}+{2\eta_{1,1}(v)-\eta_{0,2}(v)\over\eta_{1,1}(u)}
-2{\eta_{2,0}(v)\over\eta_{2,0}(u)}\right)\cdot
\mathfrak{p}_2{\p^2\over\p\mathfrak{p}_1^2}+\nn\\
&+&{1\over 2}\left({\eta_{2,0}(v)\eta_{-1,1}(qu)\over\eta_{2,0}(u)}-{\eta_{0,2}(v)\over\eta_{1,1}(u)}\right)\cdot\mathfrak{p}_1^2{\p\over\p
\mathfrak{p}_2}-\nn\\
&-&{1\over 2}\left({\eta_{2,0}(v)\eta_{-1,1}(qu)\over\eta_{2,0}(u)}-{\eta_{0,2}(v)\over\eta_{1,1}(u)}-2{\eta_{2,0}(v)\over\eta_{2,0}(u)}\right)\cdot\mathfrak{p}_2{\p\over\p
\mathfrak{p}_2}+\ldots
\ee
In such a way, one can restore the operator $\hat H(v,u)$ in terms of $\mathfrak{p}_k$-variables up to any level. An approach to a general description of these Hamiltonians can be found in \cite{GZ2}.

The only subtlety is the commutativity of $\hat H(v,u)$ at distinct values of $u$ and $v$: one can check the commutativity of (\ref{eq:19}) at any level independently due to the block-diagonal structure of the matrix, but the commutativity of (\ref{Hp}) is achieved only when acting on the space of polynomials at most of grading 2 (grading of $p_k$ is $k$), for higher gradings one needs more terms in the Hamiltonian.

Similar calculation for the $N=3$ case gives up to the second level
\be
\hat{\mathcal{O}}^{(3)}(v)\Big\{m_{[1]},m_{[1,1]},m_{[2]}\Big\}=\hat{\mathcal{O}}^{(2)}(v)\Big\{m_{[1]},m_{[1,1]},m_{[2]}\Big\}
\ee
At the first level,
\be
\hat{\mathcal{O}}^{(1)}(v)m_{[1]}=\hat{\mathcal{O}}^{(2)}(v)m_{[1]}=\hat{\mathcal{O}}^{(3)}(v)m_{[1]}
={\theta_{\tau}(v-i\hbar)\over\theta_{\tau}(v)}m_{[1]}
\ee

In \cite{MMZ}, we conjectured that
\begin{equation}
  \label{eq:23}
  \hat H(v,u) \, P_\lambda\{\mathfrak{p}_k;\hbar,g\} =\prod_{j=1}^N
  \frac{\theta_{\tau}\Big(v -i\hbar\lambda_j+(1-j)g\Big)}{\theta_{\tau}\Big(v+(1-j)g\Big)}
  \frac{\theta_{\tau}\Big(u+(1-j)g\Big)}{\theta_{\tau}\Big(u-i\hbar\lambda_i+(1-j)g\Big)}\cdot P_\lambda\{\mathfrak{p}_k;\hbar,g\}
\end{equation}
with $\mathfrak{p}_k=\sum_je^{kq_j}$. One can easily check that the polynomials $P_\lambda\{\mathfrak{p}_k;\hbar,g\}$ at $|\lambda|\le 2$ are eigenfunctions of the Hamiltonian (\ref{Hp}) with the eigenvalues as in (\ref{eq:23}).

The dual Hamiltonian that acts similarly on $E_R$ is obtained by the replaces $\mathfrak{p}_k\leftrightarrow -k\p /\p \mathfrak{p}_k$ acting to the left and $g\leftrightarrow -i\hbar$. For instance, the Hamiltonian (\ref{Hp}) becomes
\be\label{HE}
\hat H_E(v,u)&=&{\bar\eta_{1,0}(v)\over\bar\eta_{1,0}(u)}\cdot\mathfrak{p}_1{\p\over\p\mathfrak{p}_1}+\nn\\
&+&{1\over 4}\left({\bar\eta_{2,0}(v)\bar\eta_{-1,1}(qu)\over\bar\eta_{2,0}(u)}+{2\bar\eta_{1,1}(v)-\bar\eta_{0,2}(v)\over\bar\eta_{1,1}(u)}
-{\bar\eta_{1,0}(v)\over\bar\eta_{1,0}(u)}\right)
\cdot\mathfrak{p}_1^2{\p^2\over\p\mathfrak{p}_1^2}-\nn\\
&+&{1\over 4}\left({\bar\eta_{2,0}(v)\bar\eta_{-1,1}(qu)\over\bar\eta_{2,0}(u)}-{\bar\eta_{0,2}(v)\over\bar\eta_{1,1}(u)}\right)\cdot\mathfrak{p}_2{\p^2\over\p\mathfrak{p}_1^2}+\nn\\
&-&{1\over 2}\left({\bar\eta_{2,0}(v)\bar\eta_{-1,1}(qu)\over\bar\eta_{2,0}(u)}+{2\bar\eta_{1,1}(v)-\bar\eta_{0,2}(v)\over\bar\eta_{1,1}(u)}
-2{\bar\eta_{2,0}(v)\over\bar\eta_{2,0}(u)}\right)\cdot\mathfrak{p}_1^2{\p\over\p
\mathfrak{p}_2}-\nn\\
&-&{1\over 2}\left({\bar\eta_{2,0}(v)\bar\eta_{-1,1}(qu)\over\bar\eta_{2,0}(u)}-{\bar\eta_{0,2}(v)\over\bar\eta_{1,1}(u)}-2{\bar\eta_{2,0}(v)\over\bar\eta_{2,0}(u)}\right)\cdot\mathfrak{p}_2{\p\over\p
\mathfrak{p}_2}+\ldots
\ee
i.e. the coefficients in front of $\mathfrak{p}_2{\p^2\over\p\mathfrak{p}_1^2}$ and $\mathfrak{p}_1^2{\p\over\p
\mathfrak{p}_2}$ exchanged and the first one is multiplied with $-2$, while the second one, with $-1/2$ (with the corresponding exchange $g\leftrightarrow -i\hbar$, which is denoted by the bars over $\eta$'s).
Thus,
\be
\boxed{
\hat H_E(v,u)\cdot E_\lambda\{\mathfrak{p}_k;\hbar,g\}=(-1)^{|\lambda|}\prod_{j=1}^N
  \frac{\theta_{\tau}\Big(v +g\lambda_j^\vee-i\hbar(1-j)\Big)}{\theta_{\tau}\Big(v-i\hbar(1-j)\Big)}
  \frac{\theta_{\tau}\Big(u-i\hbar(1-j)\Big)}{\theta_{\tau}\Big(u+g\lambda_j^\vee-i\hbar(1-j)\Big)}\cdot E_\lambda\{\mathfrak{p}_k;\hbar,g\}
}
\ee
Hence, this way we obtain an explicit expression for the dual elliptic RS $N$-body Hamiltonians in terms of $\mathfrak{p}_k$-variables. Unfortunately, obtaining them in terms of coordinates is not immediate.

\subsection{Towards DELL Hamiltonians}

In order to construct DELL Hamiltonians, one could seem to use again the KS Hamiltonians similarly to the previous subsection. Indeed, the {\it bi}-elliptic (ell$_{p}$-ell$_{q}$) KS Hamiltonians are known both in the form (\ref{KS}) with
\be\label{Oop2}
\hat{\mathfrak{O}}(z|q_i|\tilde{\mathfrak{q}},\mathfrak{q},\hbar,g)&=&\sum_{k\in\mathbb{Z}}\hat{\mathfrak{O}}_k(q_i|\tilde{\mathfrak{q}},\mathfrak{q},\hbar,g)z^k:=\nn\\
&=&\sum_{k_1,\ldots,k_N\in\mathbb{Z}}
z^{\sum_j k_j}\mathfrak{q}^{\sum_l k_l(k_l-1)/2}\prod_{l<m}\theta\Big({g(k_l-k_m)+q_l-q_m\over 2},\tilde\tau\Big)
\prod_{j=1}^ne^{-i\hbar k_j\p_{q_j}}\nn\\
\ee
where $\tilde{\mathfrak{q}}=e^{2\pi i\tilde\tau}$ is an elliptic parameter parameterizing $\theta$-function in this formula, and in the form (\ref{eq:18}) with \cite{GZ}
\be\label{Oop3}
{\cal O}(\lambda,v)&=&
\hat{\mathfrak{O}}(\lambda,e^{v}|q_i|\tilde{\mathfrak{q}},\mathfrak{q},\hbar,g):=\nn\\
&=&\sum_{k_1,\ldots,k_N\in\mathbb{Z}}{\theta\Big(\lambda-{g\sum_ik_i\over 2},\tilde\tau\Big)\over
\theta\Big(\lambda,\tilde\tau\Big)}
e^{v\sum_j k_j}\mathfrak{q}^{\sum_l k_l(k_l-1)/2}\prod_{l<m}\theta\Big({g(k_l-k_m)+q_l-q_m\over 2},\tilde\tau\Big)
\prod_{j=1}^ne^{-i\hbar k_j\p_{q_j}}=\nn\\
&=&{1\over\det_{_{1\le k,l\le N}}\theta_k\Big(\lambda-{Nq_l-\sum_jq_j\over 2}\Big)}\cdot
\det_{{1\le k,l\le N}}\sum_ne^{nv}\mathfrak{q}^{n(n-1)/2}\theta_k\Big(\lambda-{Nq_l+Nng-\sum_jq_j\over 2}\Big)\nn\\
\ee
where $\theta_k(z)$ is the $\theta$-function with characteristics (and the modular parameter $N\tilde\tau$),
\be
\theta_k(z):=\sum_{j\in\mathbb{Z}}\exp\left[\pi i\Big(j+{1\over 2}-{k\over N}\Big)^2N\tilde\tau+\pi i\Big(j+{1\over 2}-{k\over N}\Big)\Big(z+{N\over 2}\Big)\right]
\ee
Notice that the generating functions (\ref{Oop2}) and (\ref{Oop3}) are defined in a slightly distinct way.

Moreover, the eigenfunctions of these Hamiltonians have been also conjectured in \cite{MMZ}: they are constructed basing on
the ELS-function defined as \cite{AKMM2}\footnote{Note that $s$ in \cite{AKMM2} is $e^s$ here.}
\beq
\mathfrak{P}_N { (x_i ; \tilde{\mathfrak{q}} \vert y_i ; \mathfrak{q} \vert \hbar,g,s)}
:= \sum_{\vec\lam} \prod_{i,j=1}^N
\frac{\mathcal{N}_{\lam^{(i)}, \lam^{(j)}}^{(j-i)} (e^g y_j/y_i \vert \hbar,s,\mathfrak{q})}
{\mathcal{N}_{\lam^{(i)}, \lam^{(j)}}^{(j-i)} (y_j/y_i \vert \hbar,s,\mathfrak{q})}
\prod_{\beta=1}^N \prod_{\alpha \geq 1} \left( \frac{\tilde{\mathfrak{q}} x_{\alpha + \beta}}{e^g x_{\alpha + \beta -1}} \right)^{\lam_\alpha^{(\beta)}},
\label{EG2}
\eeq
where

\beqa \label{EGfactor}
\mathcal{N}_{\lam, \mu}^{(k)} (u \vert \hbar, s, \mathfrak{q})
= \!\!\!\!\!\!\!\!
\prod_{j \geq i \geq 1 \atop j - i \equiv k~(\mathrm{mod}~n)}
\!\!\!\!\!\!\!\!
\Theta(ue^{i\hbar(\mu_i + \lam_{j+1})+s(j-i)} ; e^{-i\hbar},\mathfrak{q})_{\lam_j - \lam_{ j+1}}
\!\!\!\!\!\!\!\!
\prod_{j \geq i \geq 1 \atop j  - i \equiv -k-1~(\mathrm{mod}~n)}
\!\!\!\!\!\!\!\!
 \Theta (ue^{-i\hbar(\lam_i - \m_j+ s(i - j -1)} ; e^{-i\hbar},\mathfrak{q})_{\m_j - \m_{j+1}}
 \nn
\eeqa
and $\Theta(z;q,w)_n$ is the elliptic Pochhammer symbol (\ref{Theta}).

\bigskip

The ELS-function is naturally related to symmetric polynomials. That is, for partition $\lambda$,
\beq\label{20}
{\bf P}^\xi_\lambda(x_i;{\mathfrak{q}},\tilde{\mathfrak{q}},s,\hbar,g):=\prod_{i=1}^nx_i^{\lambda_i}\cdot
\mathfrak{P}_n^{\xi} { \left(\tilde{\mathfrak{q}}^{n-i}x_i ; \tilde{\mathfrak{q}}\, \Big|\, y_i=e^{-i\hbar\lambda_i+(s+g)(n-i)} ; \mathfrak{q} \,\Big|
\, \hbar,-i\hbar-g,s\,\right)}
\eeq
is a graded function of variables $x_i$ of the weight $|\lambda|$,
which is a series in $\tilde{\mathfrak{q}}^{nk}$:
\be\label{beP}
{\bf P}^\xi_\lambda(x_i;{\mathfrak{q}},\tilde{\mathfrak{q}},s,\hbar,g)=\sum_{k\ge 0}\tilde{\mathfrak{q}}^{nk}\cdot{{\bf P^\xi}^{(k)}_\lambda(x_i;{\mathfrak{q}},s,\hbar,g)\over \prod_{i=1}^n x_i^k}=
\sum_{k\ge 0}{\bf P^\xi}^{(k)}_\lambda(x_i;{\mathfrak{q}}s,\hbar,g)\cdot\prod_{i=1}^n\left({\tilde{\mathfrak{q}}\over x_i}\right)^k
\ee
Here ${\bf P^\xi}^{(k)}_\lambda(x_i;{\mathfrak{q}},s,\hbar,g)$ is a symmetric polynomial of variables $x_i$ with grade $|\lambda|+nk$. What is important, these polynomials do not form a complete basis at the level $|\lambda|+nk$ at $k\ne 0$. The polynomials ${\bf P^\xi}^{(0)}_\lambda(x_i;{\mathfrak{q}}s,\hbar,g)$ coincide with $P_\lambda(x_i;\hbar,g)$.

Then, the eigenfunctions of the bi-elliptic KS Hamiltonians are given, in accordance with the conjecture of \cite{MMZ},
by the limit $s\to 0$ of the ${\bf P}^\xi_\lambda(x_i;p,s,q,t)$-functions,
\begin{equation}\label{ELSKS}
\boxed{
\Psi_\mu^{KS}(x_i;{\mathfrak{q}},\tilde{\mathfrak{q}},\hbar,g):=\lim_{s\to 0}
{{\bf P}^\xi_\lambda(x_i;{\mathfrak{q}},\tilde{\mathfrak{q}},s,\hbar,g)\over \alpha^f( \tilde{\mathfrak{q}}  \vert y_i; s \vert \hbar,-i\hbar-g,\mathfrak{q})}
}
\end{equation}
with some normalization constant $\alpha^f( \tilde{\mathfrak{q}}  \vert y_i; s \vert \hbar,-i\hbar-g,\mathfrak{q})$ that makes the expression non-singular in the $s\to 0$ limit. These eigenfunctions are no longer polynomials, which is not surprising: eigenfunctions of the both elliptic Calogero and Ruijsenaars systems are not polynomials. At the same time, the ELS-functions ${\bf P}^\xi_\lambda(x_i;{\mathfrak{q}},\tilde{\mathfrak{q}},s,\hbar,g)$ are proposed to solve non-stationary equations with the KS Hamiltonians.

Thus, $\Psi_\mu^{KS}(x_i;{\mathfrak{q}},\tilde{\mathfrak{q}},\hbar,g)$ provide a bi-elliptic counterpart of the $P_\lambda\{\mathfrak{p}_k;\hbar,g\}$ polynomials. However, the problem is in constructing of what should be a generalization of the Schur scalar product that could give rise to a set bi-orthogonal to these symmetric functions $\Psi_\mu^{KS}(x_i;{\mathfrak{q}},\tilde{\mathfrak{q}},\hbar,g)$. This bi-orthogonal set would provide eigenfunctions of the DELL Hamiltonians generalizing $E_\lambda\{\mathfrak{p}_k;\hbar,g\}$ polynomials to the DELL case. Unfortunately, a substitute of the Schur scalar product for the function of the form (\ref{beP}) is unclear at the moment.

\section{Comments and discussion}

In fact, as we already noted in the Introduction, the symmetric functions ${\bf P}^\xi_\lambda(x_i;{\mathfrak{q}},\tilde{\mathfrak{q}},s,\hbar,g)$ before taking the $s\to 0$ limit are under better control as was first realized by J. Shiraishi \cite{Shi}: they are associated with non-stationary bi-elliptic system. Hence, in order to construct the DELL eigenfunctions, one would probably better deal with ${\bf P}^\xi_\lambda(x_i;{\mathfrak{q}},\tilde{\mathfrak{q}},s,\hbar,g)$ instead of  $\Psi_\mu^{KS}(x_i;{\mathfrak{q}},\tilde{\mathfrak{q}},\hbar,g)$ (i.e. {\it before} the $s\to 0$ limit is taken) and look for the corresponding lift of $E_\lambda\{\mathfrak{p}_k;\hbar,g\}$ to symmetric functions depending on two elliptic parameters {\it and} the parameter $s$.

Let us say a few words about emerging this kind of non-stationary equations. As we already emphasized, solutions to such equations are associated with (Seiberg-Witten) supersymmetric gauge theories (in $4d$ and $5d$) with codimension two defects. Using the AGT correspondence \cite{AGT,wAGT}, one can also associate them with conformal blocks in $2d$ conformal field theory with an insertion of a degenerate field \cite{MT,MMM}.

Consider, for instance, the simplest 4-point spherical conformal block with one degenerate field. It depends only on the cross-ration of the four points $x$, and satisfies the  BPZ equation \cite{BPZ}, which is just a differential equation w.r.t. $x$.
If the block is 5-point spherical, it depends on one extra parameter, and
then the equation contains also a derivative with respect to this parameter.
Similarly, the two-point toric conformal block with one degenerate field satisfies a differential equation w.r.t. the difference of positions of these two fields {\it and} w.r.t. the torus modular parameter $\tau$. The terminology refers to the simplest example of the field, which is degenerate at the second level: then, in these both cases, the equations looks like non-stationary Schr\"odinger equations. On the integrable theory side, the 5-point spherical conformal block gives the $XXX$ spin chain equation and the 2-point toric one, the non-stationary Schr\"odinger equation with the elliptic Calogero-Moser Hamiltonian \cite{MT,MMM}.

If one considers instead of the Virasoro algebra the $q$-Virasoro algebra \cite{qVir},
the equations get modified and can become difference rather than differential, and the degenerate 5-point spherical conformal block has to correspond to the $XXZ$ spin chain equation \cite{Sham} (see also \cite{ASh}), while the 2-point toric one, to the non-stationary Schr\"odinger equation with the elliptic RS Hamiltonian \cite{Shi}. In this paper, we were interested in the latter case. Similarly, further deformation of the $q$-Virasoro algebra to an elliptic Virasoro algebra would give rise to the non-stationary DELL equation satisfied by the corresponding 2-point toric conformal block with a degenerate field inserted. This conformal block, in the limit ${\mathfrak{q}}\to 0$, gives the non-stationary RS wave function (Shiraishi series) \cite{Shi}, in the further limit $s\to 0$, the eigenfunctions of the elliptic RS Hamiltonians, and, in the limit $s,\tilde{\mathfrak{q}}\to 0$, gives the $E_\lambda$-polynomials.

\newpage
At this point, we can understand that there are two algebras hidden behind the system. On one hand, say, we deal with the 2-point torus conformal block (with a degenerate field) of the $q$-Virasoro algebra, which is built from generators of the Ding-Iohara-Miki (DIM) algebra \cite{DIM,AF,AKMMMZ}. On the other hand,  this conformal block is a lift to $s\ne 0$ (non-stationarity) of the elliptic RS eigenfunctions, i.e. of the $E_\lambda$ polynomials considered as functions of $\lambda$. In its turn, the $E_\lambda$ polynomials are associated \cite{MMZ} with the elliptic DIM algebra \cite{elDIM}, i.e. with the DIM algebra \cite{DIM} with one direction compactified \cite{MMZ1}  (or, equivalently, with the trigonometric DIM algebra with an additional Heisenberg subalgebra \cite{elDIMZ}). That is, the vertical Heisenberg subalgebra of the elliptic DIM algebra provides the Hamiltonians dual to the elliptic RS, whose eigenfunctions are $E_\lambda$ polynomials \cite{MMZ}. Hence, in this case, there are two algebras underlying the system: the ordinary trigonometric DIM and the elliptic DIM algebras.

Similarly, the elliptic Virasoro algebra is presumably built from the elliptic DIM algebra in the same way as the $q$-Virasoro algebra is obtained from the trigonometric DIM algebra \cite{AKMMMZ}, and its 2-point toric conformal block with a degenerate field is a non-stationary DELL wave function, while the DELL Hamiltonians are associated with
the double elliptic DIM algebra (so called Pagoda algebra), i.e. the DIM algebra with two directions compactified  \cite{MMZ1}.

On the physical side, the three cases (Virasoro, $q$-Virasoro, elliptic Virasoro) are associated with $4d$, $5d$ and $6d$ supersymmetric theories with adjoint matter hypermultiplet, and the 2-point toric conformal block is associated with the codimension two defect inserted. The parameters that emerged, e.g., in ${\bf P}^\xi_\lambda(x_i;{\mathfrak{q}},\tilde{\mathfrak{q}},s,\hbar,g)$, are associated with parameters of the gauge in the following way \cite{AKMM2}:
\begin{itemize}
\item[$\hbar$, $s$] are two $\Omega$-background deformation parameters on the gauge theory side, $\hbar=2\pi\epsilon_1$, $s=-2\pi\epsilon_2$, the limit of $s\to 0$ reduces the system to the quantum integrable system, and a (hypothetical) non-stationary DELL equation to the eigenvalue DELL Hamiltonian problem. On the Virasoro side, they rescale the dimensions of the operators.
\item[$g$] is the central charge parameter on the algebra side, and the coupling constant parameter on the integrable side. On the gauge theory side, it is related to the mass of the adjoint hypermultiplet, $g=-m$.
\item[$\tilde{\mathfrak{q}}$] is the elliptic parameter that controls the coupling in the gauge theory (the bare torus and the bare charge). On the integrable side, it is associated with the torus where the coordinates live. On the Virasoro side, it is associated with the torus where the $2d$ fields in the $4d$ limit live.
\item[$\mathfrak{q}$] is the elliptic parameter that is associated with the Kaluza-Klein torus in the gauge theory (one considers Seiberg-Witten (SW) $6d$ theory with two dimensions compactified onto a $2d$ torus). On the integrable side, it is associated with the torus where the momenta live.
\end{itemize}

These correspondences are collected in the following table.

\bigskip

\hspace{-.8cm}\begin{tikzpicture}
\draw (0,0) rectangle (17.5,9);
\draw (0,4) -- (17.5,4);
\draw (0,2) -- (17.5,2);
\draw (0,6) -- (17.5,6);
\draw (2.5,0) -- (2.5,9);
\draw (5,0) -- (5,9);
\draw (7.5,0) -- (7.5,9);
\draw (10,0) -- (10,9);
\draw (12.5,0) -- (12.5,9);
\draw (15,0) -- (15,9);
\draw (1.25,8.5) node{};
\draw (1.25,7.5) node{SW theory};
\draw (1.25,6.5) node{};
\draw (3.75,8.5) node{Conformal};
\draw (3.75,7.5) node{block of};
\draw (3.75,6.5) node{algebra $G$};
\draw (6.25,8.6) node{Double loop};
\draw (6.25,7.9) node{algebra};
\draw (6.25,7.2) node{underlying};
\draw (6.25,6.5) node{algebra $G$};
\draw (8.75,8.7) node{Integrable};
\draw (8.75,8.1) node{system};
\draw (8.75,7.5) node{associated};
\draw (8.75,6.9) node{with toric};
\draw (8.75,6.3) node{conf. block};
\draw (11.25,8.7) node{Double loop};
\draw (11.25,8.1) node{algebra};
\draw (11.25,7.5) node{containing};
\draw (11.25,6.9) node{dual integrable};
\draw (11.25,6.3) node{Hamiltonians};
\draw (13.75,8.7) node{Integrable};
\draw (13.75,8.1) node{system};
\draw (13.75,7.5) node{associated};
\draw (13.75,6.9) node{with spherical};
\draw (13.75,6.3) node{conf. block};
\draw (16.25,8.7) node{Integrable};
\draw (16.25,8.1) node{system};
\draw (16.25,7.5) node{associated};
\draw (16.25,6.9) node{with the pure};
\draw (16.25,6.3) node{gauge limit};
\draw (1.25,5) node{$4d$};
\draw (1.25,3) node{$5d$};
\draw (1.25,1) node{$6d$};
\draw (3.75,5) node{$G=$Virasoro};
\draw (3.75,3) node{$G=q$-Virasoro};
\draw (3.75,1.4) node{$G=$elliptic};
\draw (3.75,.7) node{Virasoro};
\draw (6.25,5) node{Affine Yangian};
\draw (6.25,3) node{DIM};
\draw (6.25,1) node{Elliptic DIM};
\draw (8.75,5.4) node{Elliptic};
\draw (8.75,4.7) node{Calogero-Moser};
\draw (8.75,3) node{Elliptic RS};
\draw (8.75,1) node{DELL};
\draw (11.25,5.4) node{Elliptic affine};
\draw (11.25,4.7) node{Yangian};
\draw (11.25,3) node{Elliptic DIM};
\draw (11.25,1.4) node{Double elliptic};
\draw (11.25,.7) node{DIM (Pagoda)};
\draw (13.75,5.4) node{periodic XXX};
\draw (13.75,4.7) node{spin chain};
\draw (13.75,3.4) node{periodic XXZ};
\draw (13.75,2.7) node{spin chain};
\draw (13.75,1.4) node{periodic XYZ};
\draw (13.75,.7) node{spin chain};
\draw (16.25,5.4) node{periodic};
\draw (16.25,4.7) node{Toda chain};
\draw (16.25,3.5) node{periodic};
\draw (16.25,3) node{relativistic};
\draw (16.25,2.5) node{Toda chain};
\draw (16.25,1.5) node{periodic};
\draw (16.25,1) node{elliptic};
\draw (16.25,.5) node{Toda chain};
\end{tikzpicture}

\bigskip

In this table, in the first column, there are Seiberg-Witten theories with adjoint matter; in the second column, there are conformal blocks of an algebra $G$, which are associated with insertion of codimension two defects on the Seiberg-Witten theory side; in the third column, there are DIM type (double loop) algebras underlying the algebra $G$; in the fourth column, there are integrable systems, non-stationary equations in these systems being solved by the toric conformal block; in the fifth column, there are algebras containing the Hamiltonians associated with duals to these integrable systems; in the sixth column, there are integrable systems associated with spherical conformal blocks of the same algebra $G$. Note that the pure gauge theory limit (irregular conformal block \cite{Gai,MMMir}) gives rise to the same integrable systems in the both cases of toric and spherical conformal blocks, this is reflected in the last, seventh column. For the association with integrable systems, see \cite{GM} and references therein.

\section{Conclusion}

To summarize, this paper provides an exposition of the present status of the old DELL puzzle
formulated in \cite{MM}.
In physical language, the main difficulty is elliptization of momenta,
and the crucial achievement, which we outline for the first time,
can be an explicit description of the dual elliptic RS system.
In mathematical language, the task is to build a complete theory of Shiraishi functions \cite{Shi,AKMM2,GNS}
and study their various limits, relevant for the duality problem.
The real (unresolved) mystery is matching these results with the only known
DELL Hamiltonian: that at $N=2$ \cite{BMMM},
which  contains a sophisticated dynamical (coordinate-dependent) mixture of the
two elliptic parameters,
i.e. describes the DELL system as a highly non-trivial bi-elliptic system,
much more involved than the straightforward Koroteev-Shakirov Hamiltonians \cite{KS}.

Another type of questions is to find a place of duality in the modern studies
of the WLLZ models \cite{Ch}, where whole {\it families} of integrable hierarchies \cite{Ch12,MMMP1} and of
classes of many-body systems \cite{Lit,MMCal,MMMP1,MMMP2} emerge
which are not so simple to $q,t$-deform \cite{GKLMM,Ch3,Bourgine}.
The theory of Shiraishi functions and DIM representations can be probably naturally treated within this context,
but what substitutes the $p,q$-duality is still somewhat unclear.
We are looking forward towards some new progress in this difficult but promising field.

\section*{Acknowledgements}

We are grateful to Yegor Zenkevich for numerous fruitful discussions of the duality problem during a few recent years. We are also indebted to Sergei Kharchev for a detailed discussion on the subject of sec.\ref{amb}.  Our work is supported by the Russian Science Foundation (Grant No. 23-41-00049).

\section*{Appendix}

\subsection*{Some properties of (\ref{N2sd})}

Let us consider the Macdonald polynomial for the symmetric Young diagram and at $N=2$ with $x_1=u^\mu v^{-1}$, $x_2= v^{-2}$:
\be
M_{[\lambda]}\Big(u^\mu v^{-1},v^{-2};u,v\Big)\stackrel{(\ref{M2})}{=}u^{\mu\lambda}v^{-\lambda}
\sum_{m=0}^\lambda \frac{1}{u^{(1+\mu)m}w^m} \prod_{j=0}^{m-1} c_{\lambda,j}(u,v)
:=u^{\mu\lambda}v^{-\lambda}M(\lambda,\mu)
\ee
Then, the duality equation (\ref{deq}) or (\ref{N2sd}) reduces to
\be
\boxed{
M(\mu,0)M(\lambda,\mu) = M(\lambda,0)M(\mu,\lambda)
}
\ee
In particular, at $v=u$ all $ c_{\lambda,j}(u,v)=1$, thus
\be
M(\lambda,\mu) \stackrel{v=u}{=} \sum_{m=0}^\lambda \frac{1}{u^{(1+\mu)m} }
= \frac{1-\frac{1}{u^{\lambda+1)(\mu+1)}}}{1 - \frac{1}{u^{\mu+1}}}
\ee
and
\be
M(\mu,0) \  \stackrel{w=1}{=}\ \frac{1 - \frac{1}{u^{\mu+1}}}{1-\frac{1}{\mu}}
\ee
so that
\be
M(\mu,0)M(\lambda,\mu)\ \stackrel{v=u}{=} \
\frac{1-\frac{1}{u^{\lambda+1)(\mu+1)}}}{1-\frac{1}{\mu}}
\  \stackrel{v=u}{=}\ M(\lambda,0)M(\mu,\lambda)
\label{w=1}
\ee

In another extreme case, $v/u=\infty$, only the term $m=0$ contributes to
\be
M(\lambda,\mu) \ \stackrel{v/u=\infty}{=} \ \frac{u^\lambda-1}{u^{\lambda-1}(u-1)}
\ee
so that $M(\lambda,\mu) \stackrel{v/u=\infty}{=} M(\lambda,0)$ and
$M(\mu,\lambda,\mu) \stackrel{v/u=\infty}{=} M(\mu,0)$.

\subsection*{Elliptic counterpart of $M(\lambda,\mu)$}

Eq.(\ref{w=1}) is actually
\be
\sin(n x) \sim \prod_{j=0}^{n-1} \sin\left(x+\frac{j}{n}\pi\right)
\sim \prod_{j=0}^{n-1} \left(1 - e^{2ix}e^{\frac{2\pi i}{n}}\right)
\sim \sin x \sum_{m=0}^{n-1} e^{2imx}
\ee
(e.g. $\sin(2x) \sim \sin(x) \sin(x+\frac{\pi}{2}) = \sin (x)\cos (x)$ etc),
which is an elementary consequence of the infinite-product formula for $\sin$.

It follows that
\be
\frac{1-e^{2(\lambda+1)ix}}{1-e^{2ix}}
= \sum_{m=0}^\lambda e^{2im x}
\ee
In (\ref{w=1}), $n=\lambda+1$ and  $e^{2ix} = u^{-(\mu+1)}$.
%
%
Its direct elliptic counterpart (also following from the infinite-product formula) is
\be
\theta_1(nx)\sim \prod_{i,j=0}^{n-1} \theta_1\left(x+\frac{i+j\tau}{n}\right)
\label{thetaMx}
\ee

Thus it ``remains" to unify this elliptic deformation of a sum in $M(\lambda,\mu)$
with an ``obvious" elliptic deformation of the product.
Probably, important for this is to convert the {\it pair} of product indices $(i,j)$
into a single summation index $m$, as above.

Hence, an elliptic counterpart of $M(\lambda,\mu)$ should have the form
\be
{\cal B}(\lambda,\mu) \  \stackrel{???}{=} \
\sum_{m=0}^\infty
{\rm res}_m\left\{
\frac{\theta_1\Big((\lambda+1)(\mu+1)\xi+\zeta |\tau\Big)}{\theta_1(\xi+\zeta)}\right\}
\  \prod_{j=0}^{m-1}
\frac{\theta_1(\lambda+1-j)\xi  |\tau')}{\theta_1((\lambda+1-j)\xi +\zeta |\tau' )}
\frac{\theta_1(((j+1)\xi  +\zeta |\tau' )}{\theta_1( (j+1)\xi  |\tau' )}
\ee
with some residues (contour integrals)
of (\ref{thetaMx}), picking up a particular item of the series in $u=e^{i\xi}$ or $v/u=e^{i\zeta}$.

\end{document}